\documentclass[fleqn,usenatbib]{mnras}

\usepackage{newtxtext,newtxmath}

\usepackage[T1]{fontenc}

\DeclareRobustCommand{\VAN}[3]{#2}
\let\VANthebibliography\thebibliography
\def\thebibliography{\DeclareRobustCommand{\VAN}[3]{##3}\VANthebibliography}

\usepackage{graphicx}	

\usepackage{amsmath}	
\usepackage{amssymb}	
\usepackage{multicol}   
\usepackage{bm}
\usepackage{color}
\usepackage{xspace}
\usepackage{CJKutf8}

\definecolor{mygray}{gray}{0.6}
\definecolor{magenta}{rgb}{0.858, 0.188, 0.478}
\newcommand{\ccc}[1]{\textcolor{red}{[\textit{\small #1}]}}

\newcommand{\hcadd}[1]{{#1}}

\newcommand{\hcch}[2]{{{#2}}}
\newcommand{\corem}[1]{\textcolor{mygray}{\sout{#1}}}
\newcommand{\hcrem}[1]{}

\newcommand{\xxx}[1]{\textcolor{blue}{\textbf{xxx}\xspace}}

\newcommand{\fg}[1]{Fig.~\ref{fig:#1}}
\newcommand{\Fg}[1]{Figure~\ref{fig:#1}}

\newcommand{\eq}[1]{Eq.~(\ref{eq:#1})\xspace}
\newcommand{\Eq}[1]{Equation~(\ref{eq:#1})\xspace}

\newcommand{\tb}[1]{Table~\ref{tab:#1}\xspace}
\newcommand{\Tb}[1]{Table~\ref{tab:#1}\xspace}
\newcommand{\se}[1]{Sect.~\ref{sec:#1}\xspace}
\newcommand{\Se}[1]{Section~\ref{sec:#1}\xspace}

\usepackage{ulem}
\makeatletter
\def\uwave{\bgroup \markoverwith{\lower3.5\p@\hbox{\sixly \textcolor{red}{\char58}}}\ULon}
\font\sixly=lasy6 
\makeatother

\renewcommand{\ccc}[1]{}
\renewcommand{\corem}[1]{}

\renewcommand{\uwave}[1]{#1}

\title[Survival of ALMA Rings]{Survival of ALMA Rings in the Absence of Pressure Maxima}

\author[Jiang \& Ormel]{
Haochang Jiang (\begin{CJK*}{UTF8}{gbsn}蒋昊昌\end{CJK*})$^{1}$, 
Chris W. Ormel$^{1}$\thanks{E-mail: chrisormel@tsinghua.edu.cn}
\\
$^{1}$Department of Astronomy, Tsinghua University, Haidian DS 100084, Beijing, China
}

\date{Accepted XXX. Received YYY; in original form ZZZ}

\pubyear{2021}

\begin{document}
\label{firstpage}
\pagerange{\pageref{firstpage}--\pageref{lastpage}}
\maketitle

\begin{abstract}
    Recent ALMA observations have revealed that a large fraction of protoplanetary discs contain bright rings at (sub)millimeter wavelengths. Dust trapping induced by pressure maxima in the gas disc is a popular explanation for these rings. However, it is unclear whether such pressure bumps can survive for evolutionary time-scales of the disc. In this work, we investigate an alternative scenario, which involves only dust-gas interactions in a smooth gas disc. We postulate that ALMA rings are a manifestation of a dense, clumpy mid-plane that is actively forming planetesimals. The clumpy medium itself hardly experiences radial drift, but clumps lose mass by disintegration and vertical transport and planetesimal formation. Starting from a seed ring, we numerically solve the transport equations to investigate the ring's survival. In general, rings move outward, due to diffusion of the clump component. Without pressure support, rings leak material at rates $\sim$40 $M_\oplus\,\mathrm{Myr}^{-1}$ and in order for rings to survive, they must feed from an external mass reservoir of pebbles. In the case where the pebble size is constant in the disk, a cycle between ring formation and dispersion emerges.
    Rings produce large quantities of planetesimals, which could be material for planet formation and explain the massive budget inferred debris disc. Mock images of ALMA observations compare well to the rings of Elias 24 and AS 209 from DSHARP's sample.
\end{abstract}




\section{Introduction}

Thanks to the high sensitivity and spatial resolution, the Atacama Large Millimeter\hcadd{/submillimeter} Array (ALMA) has provided us with unprecedented imagery starting with HL tau \citep{ALMAPartnershipEtal2015}. At (sub)millimeter wavelengths, these images reveal the disc mid-plane materials, which harbors most of the discs' mass. At continuum bands, the emission originates from so-called pebble-sized particles -- particles of moderate size, significantly larger than micron-size ISM dust grains but small enough to provide an opacity at $\sim$mm wavelengths. These are the precursors of planetary objects: planetesimals and the cores of giant planets.

Intriguingly, ALMA found that these planet-building blocks are not homogeneously distributed, contrasting decades-long beliefs that disks are ``smooth'' \citep{Weidenschilling1977i,Hayashi1981}. Instead, disks manifest structure in diverse forms, ranging from spirals, gaps, lopsided arcs to rings. Particularly abundant are the axisymmetric rings, which appear in at least 17 out of 18 samples of the DSHARP survey \citep{HuangEtal2018}. Even though the DSHARP survey is biased towards bright sources, these annular substructures are hosted by stars across a range of luminosities, masses, and accretion rates \citep{ZhangEtal2016,AndrewsEtal2018,LongEtal2018,vanderMarelEtal2019,CiezaEtal2019,FrancisvanderMarel2020}. The morphology of the substructure shows remarkable diversity as well. Rings can appears at almost any radius (5$-$150 au) and their widths range from a few astronomical units to tens of astronomical units even in the disc. This observational diversity perhaps suggests that the origins of substructures may also be various \citep{Andrews2020}.

There are many hypotheses aiming to explain the observed substructures. Arcs may reveal that pebbles are concentrated by vortices, while spirals or other non-axisymmetric structure may hint at the presence of gravitationally unstable discs \citep{Toomre1964,Boss1997,KratterLodato2016} or massive companions \citep{DongEtal2015,ZhuEtal2015i,BaeZhu2018}. However, nonaxisymmetric substructures are relatively rare \citep{Andrews2020}. In this work, we focus on the most common substructure -- rings. Explanation of substructure formation can be grouped in two camps: with and without pressure maxima.

\hcch{In t}{T}he iceline scenario \hcadd{is one possible mechanism where }\hcch{, a}{the} pressure bump is \hcch{ needed}{not needed for ring formation}. The iceline explanation exploits that the mechanical properties of particles depend on the composition of the surface material which makes up the dust grains. For example, ice mantles can evaporate off grains when particles are exposed to higher ambient temperatures \citep[e.g.,][]{ZhangEtal2015,PinillaEtal2017} or the particles become more brittle due to sintering  \citep{OkuzumiEtal2016,SironoUeno2017}. This results in a radial dependence of --say-- the fragmentation threshold, which in turn changes the physical characteristics (the particle size), their concentration (smaller solids drift slower and pileup), and hence the appearance. Therefore, the particles size can change rapidly after they cross condensation fronts. Various hydrodynamic models caused by gas–particle coupling can also generate annular substructure, for example, self-induced pileup of particles by aerodynamical feedback \citep{DrazkowskaEtal2016,GonzalezEtal2017}. When gas drag slows down the solids' self-gravitational collapse, secular gravitational instability (SGI) happens and breaks down the over-densities into narrow rings \citep[e.g.,][]{TakahashiInutsuka2014,TominagaEtal2020}.

A more direct explanation is particles trapping due to a reversal of the pressure gradient \citep{KretkeLin2007,PinillaEtal2012,DullemondEtal2018}. In smooth disks, the pressure radially decreases from the hot and dense inner disk to the cool and more tenuous outer disk. This pressure gradient provides hydrostatic support, causing the disk to rotate at sub-Keplerian velocity and causing particles to drift inwards \citep{Weidenschilling1977,BrauerEtal2007}. A local reversal of the pressure gradient, however, would result in the opposite behaviour: gas would rotate super-Keplerian and particles would drift outwards. Hence, particles pile up at pressure maxima. Applied to ALMA rings, the idea is that rings are associated with these pressure maxima. The boundary of magnetically dead and active zones is one possible site to set such a pressure bump for ring formation \citep[e.g.,][]{FlockEtal2015,PinillaEtal2016}. MHD zonal flows, creating narrow enhancements in pressure, are expected to concentrate solids \citep[e.g.,][]{UribeEtal2011,BaiStone2014,SurianoEtal2017,SurianoEtal2018}. A more popular explanation are disc-planet gravitational interactions. As planets gravitational push aside the gas in the disc \citep{LinPapaloizou1986}, they are naturally considered to be the ``unseen'' driving forces behind pressure bumps \citep[e.g.,][]{RiceEtal2006,PaardekooperMellema2006,ZhuEtal2012,DipierroEtal2016,BaeEtal2017}. In its vicinity, the planet changes the rotation profile in a non-axisymmetric fashion. When this kinematic signal is observed \citep{TeagueEtal2018,PinteEtal2020} the case for the planet explanation is particularly compelling. 

There are several reasons why it is unlikely that all rings are associated with planets. First, planets need to be of appreciable mass to perturb the disk, unless the disk is extremely inviscid \citep{RosottiEtal2016,DongEtal2017}. However, from GPI exoplanet survey and SPHERE infrared survey for exoplanets (SHINE), it is known that wide-orbit ($>$5 au) giant planets are rare \citep[occurrence $\lesssim$5$\%$ for FGK stars,][]{NielsenEtal2019,ViganEtal2020}. Another problem is that rings, due to their ubiquity, need to last for evolutionary times. In other words, how do rings survive for million years. Specifically, a planet large enough to open a gap in the gas disk, is also in its migration "sweet spot" (between the Type I and Type II migration regimes) and as such also tends to migrate on time-scales shorter than the evolutionary times. In addition, the pileup of material in the ring region should provide the optimal conditions of ongoing planet formation -- a high density of material in a pressure bump -- begging the question how this material just sits around for $\sim$Myr. Finally, blocking all pebbles at large distance is clearly unfavorable for the ability to form planets in the interior disc, conflicting the high occurrence of close-in planets inferred from the exoplanet census \citep[e.g.,][]{ZhuEtal2018}.

In this work, we investigate whether the survival of rings can yet be reconciled by a model where the gas disk is none the less smooth. More precisely, we focus \textit{not} on the formation of the annular structure, which may well be due to one of the afore mentioned ring formation processes, like a pressure bump, but could also be attributed to the star formation process \citep[e.g.,][]{BateEtal2010}, but rather on its long-term survival in a disk characterized by a "standard" smooth pressure gradient. In such setups rings will leak a great amount of material and, in order to survive, must be replenished by an external reservoir of pebbles. Our goal is to quantify the conditions that must be fulfilled in order to match the ALMA and DSHARP observations, and to put the implications of our ring model in the context of planet formation.

Particles clumping has been observed in many shearing-box simulations studying the streaming instability \citep[e.g.,][]{JohansenYoudin2007,BaiStone2010}. Systematic studies on clumping in stratified discs reveal that axisymmetric pebble filaments form via streaming instabilities, before they collapse into planetesimals \citep{YangJohansen2014,YangEtal2017,LiEtal2018}. Motivated by the pebble clumping, we conceive a scenario where solids are divided into two components: freely-drifting "disc pebbles" and nearly-stationary (mid-plane) "clump pebbles". In our simulations, a seed ring is initialized at 10--100\,au. This perturbation will grow by absorbing drifting pebbles at its leading edge, but also lose a great number of pebbles at its back as there is no pressure confinement. In the clumpy ring model (hereafter CRM), planetesimal formation is also accounted for and plays a crucial role during the rings evolution. By numerically solving the 1D mass transport equations, we assess the conditions for which a ring-like feature can be sustained over several million years. 

The plan of the paper is as follows. The dust transport models are described in \Se{model}. The numerical results are presented in \Se{results}. We apply this model to DSHARP samples and compare with observations in \Se{ALMA}. \Se{discussion} presents an assessment of the model and outlines implications for planet formation. We summarize the main results and conclusions in \Se{conclusions}.

\section{Model}\label{sec:model}

\begin{figure*}
\centering
\includegraphics[width=\textwidth]{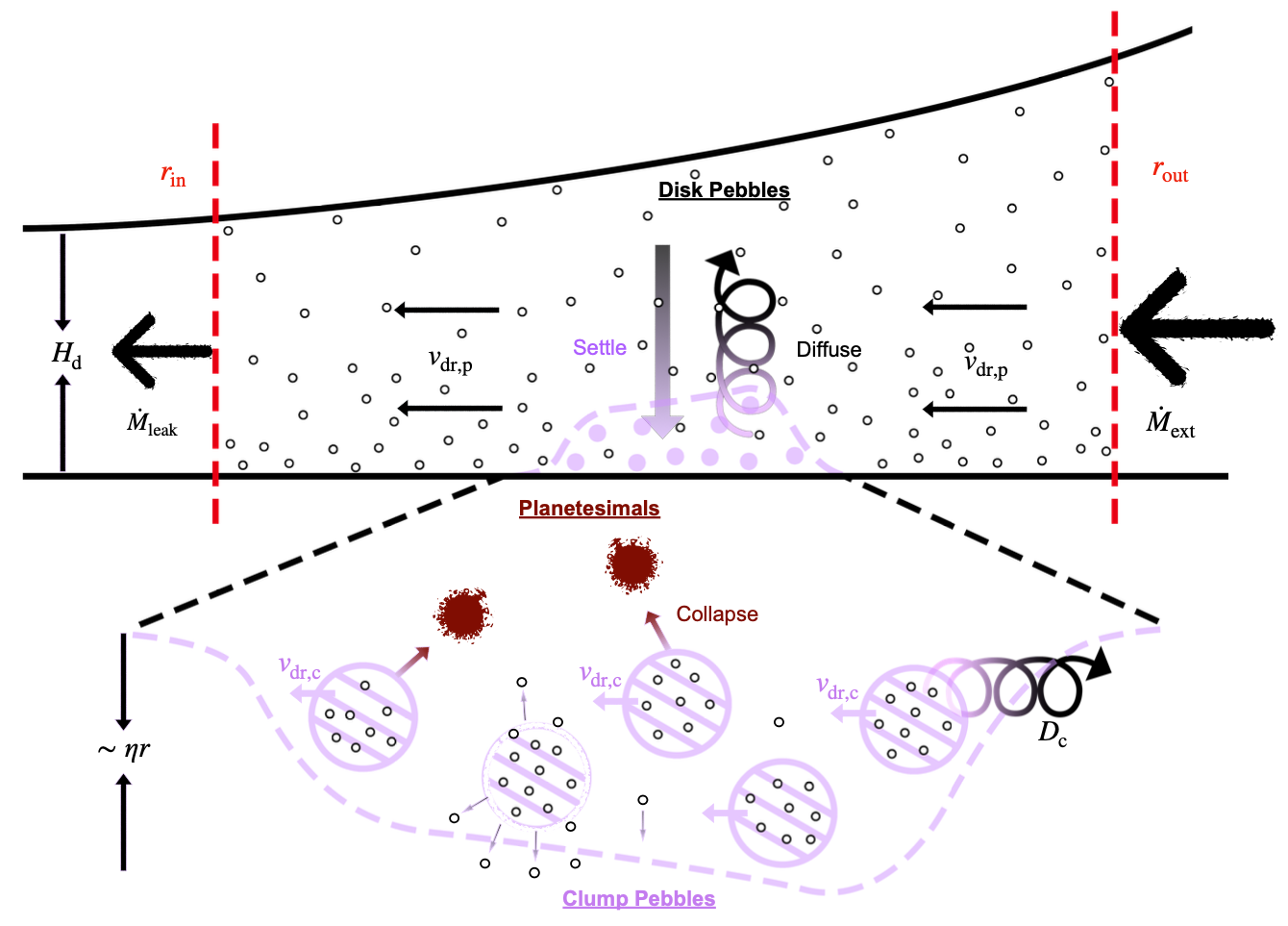}
\caption{Sketch of the model in this paper. The three components are separately shown in black (disc pebbles), purple (clump pebbles; the ring) and brown (planetesimals). A constant external mass flux $\dot{M}_\mathrm{ext}$ is imposed on the outer boundary $r_\mathrm{out}$ where disc pebbles drift in at a velocity $v_\mathrm{dr,p}$. Disc and clump pebbles continually interchange material by settling (followed by their incorporation into clumps) and clump dispersal (followed by vertical diffusion). Clump pebbles inside the ring hardly drift toward the central star ($v_\mathrm{dr,c} \ll v_\mathrm{dr,p}$) and can collapse into planetesimals, which in the CRM do not interact with the other components. The clump layer has a thickness $\sim\!\eta r$, and the clump particles are characterized by a low but non-zero radial diffusivity ($D_\mathrm{c}$) due to clump dispersal and vertical diffusion. A leaking mass flux $\dot{M}_\mathrm{leak}$ is released from the ring.}
\label{fig:cartoon}
\end{figure*}

In the CRM, it is assumed that a single particle size dominates the mass budget of the pebbles. We do not consider a particle size distribution. Solids are divided among three components:
\begin{enumerate}
    \item Disc pebbles. Dust that is vertically extended, homogeneously distributed, and which drifts inwards.
    \item Clump pebbles. Dust that resides in a clumpy state in the mid-plane regions. Clumps hardly drift, but are subject to radial diffusion.
    \item Planetesimals. Km-sized particles formed after clumps have experienced gravitational collapse. They do not interact with the other two components.
\end{enumerate}
A cartoon of the CRM is shown in \fg{cartoon}. The disc gas surface density profile has no substructure and follows a standard power-law. Disc pebbles and clump pebbles continually interchange material vertically by settling and diffusion. Clump pebbles collapse into planetesimals. For simplicity in this paper, planetesimals do not interact with the other components. We treat disc pebbles as vertically extended, at a scale height no larger than the gas scale height. As the dust-to-gas ratio of disc pebble increases, the settling of disc pebble creates a new sublayer in the protoplanet disc \citep{Sekiya1998,YoudinShu2002}. Within this layer the local dust-to-gas ratio exceeds unity; it rotates Keplerian and pebbles trapped in this layer hardly drift toward the central star. Clumps yet diffuse radially. Clumps either disperse to return pebbles back to the disc component or collapse to trigger planetesimal formation.

\subsection{Transport Model}\label{sec:transportmodel}
We consider the time evolution of the surface density of disc dust and clump dust in $1$D cylindrical geometry. The transport equation for the two pebble components (disc pebbles and clump pebbles) are
\begin{equation}
\begin{aligned}
\label{eq:transport}
    \frac{\partial \Sigma_\mathrm{d}}{\partial t} + \frac{1}{r}\frac{\partial}{\partial r}\left(r \Sigma_\mathrm{d} v_\mathrm{dr,d}\right) - \frac{1}{r}\frac{\partial}{\partial r}\left[r D_\mathrm{d} \Sigma_\mathrm{g} \frac{\partial}{\partial r}\left(\frac{\Sigma_\mathrm{d}}{\Sigma_\mathrm{g}}\right)\right] &= S_\mathrm{cp}
    \\
    \frac{\partial \Sigma_\mathrm{c}}{\partial t} + \frac{1}{r}\frac{\partial}{\partial r}\left(r \Sigma_\mathrm{c} v_\mathrm{dr,c}\right) - \frac{1}{r}\frac{\partial}{\partial r}\left[r D_\mathrm{c} \Sigma_\mathrm{g} \frac{\partial}{\partial r}\left(\frac{\Sigma_\mathrm{c}}{\Sigma_\mathrm{g}}\right)\right] &= -S_\mathrm{cp} - S_\mathrm{plt}
\end{aligned}
\end{equation}
where $\Sigma_\mathrm{g}$, $\Sigma_\mathrm{d}$, $\Sigma_\mathrm{c}$ are, respectively, the gas, disc pebble, and clump pebble surface densities. \hcch{T}{In analogy to the \citet{ShakuraSunyaev1973}'s parametrization for the shear stress, t}he turbulent particle diffusivity of the clump pebbles is \hcadd{defined as} $D_\mathrm{c} = \alpha_\mathrm{c} \Omega_K H_\mathrm{g}^2$ \hcrem{with $\alpha_\mathrm{c}$ the usual parameterization} \hcadd{\citep{YoudinLithwick2007,YangEtal2018}, where $\Omega_K(r) = \sqrt{GM_\star/r^3}$ is the local Keplerian frequency with host stellar mass $M_\star$}. Similarly, for disc pebble it is $D_\mathrm{d} = \alpha_\mathrm{d} \Omega_K H_\mathrm{g}^2$. As the radial transport of disc pebble is dominated by advection, we omit its radial diffusion in most of our simulations ($\alpha_\mathrm{d} = 0$), but see \se{ringsuv} for the non-zero $\alpha_\mathrm{d}$ cases.

The radial drift velocity is \citep[e.g.][hereafter referred to as the NSH solution]{NakagawaEtal1986}
\begin{equation}\label{eq:vdr}
    v_\mathrm{dr} = \frac{\mathrm{St}}{(1+Z)^2+\mathrm{St}^2} \dfrac{c_s^2}{\Omega_Kr}\dfrac{\partial \log{P}}{\partial \log{r}} = -\frac{2\mathrm{St}}{(1+Z)^2+\mathrm{St}^2} \eta v_K
\end{equation}
where $v_K = \Omega_K r$ is the Keplerian velocity, $Z$ is the local dust-to-gas mass ratio \hcadd{of each component}, and $\eta \equiv -(c_s^2/2v_K^2)(\partial \log{P}/\partial \log{r})$ is a dimensionless measure of the radial pressure gradient. \hcadd{For disc pebble and clump pebble, $Z$ takes on different values according to their surface densities and scale heights (see \se{discmodel} and \se{clumpmodel}). The drift velocity of disk pebbles is always given by \eq{vdr}. However, for clump pebbles we consider several options for their drift motion:}
\begin{enumerate}
    \item Clumps drift at the NSH solution
    \item Clumps drift faster than the NSH solution
    \item Clumps do not drift at all, $v_\mathrm{dr}=0$
\end{enumerate}
In the first two cases the drift velocity is determined by the $f_\mathrm{ca}$ parameter
\begin{equation}\label{eq:fca}
    v_\mathrm{dr,c} = f_\mathrm{ca} v_\mathrm{dr}(Z) + (1-f_\mathrm{ca})v_\mathrm{idv}
\end{equation}
that is, a mixture between the individual velocity ($f_\mathrm{ca}=0$)\hcadd{, where $v_\mathrm{idv} = v_\mathrm{dr}(Z=0)$,} and the NSH solution ($f_\mathrm{ca}=1$).
The value of $f_\mathrm{ca}$ that we used is an input parameter to the CRM. It is listed in \Tb{models}. When it is absent, the clump advection velocity used in \eq{transport} is $0$ (case iii). 

A clump velocity exceeding the NSH solution is motivated by the simulations of \citet{JohansenYoudin2007}, where it was found that radial drift can be faster than the NSH solution. However, these simulations were conducted without vertical gravity. On the other hand, \citet{YangJohansen2014} and \citet{YangEtal2017}, accounting for vertical gravity, found that the regions with above-average pebble concentrations (a.k.a.\ clumps), were characterized by a virtually-zero clump drift.

The RHS of \Eq{transport} are source terms. The material exchange between mid-plane and upper layer disc is
\begin{equation}\label{eq:sourcecd}
  S_\mathrm{cp} = f_\mathrm{\tau}\frac{\Sigma_\mathrm{c}}{t_\mathrm{diff}}-f_\mathrm{c}\frac{\Sigma_\mathrm{d}}{t_\mathrm{sett}}.
\end{equation}
The first term on the right of \Eq{sourcecd} is the material lost from clumps due to turbulent diffusion and the second term describes the gain of clumps from disc pebbles that settle to the mid-plane. There are two modulation factors, $f_\mathrm{\tau}$ and $f_\mathrm{c}$, with values between 0 and 1. The factor $f_\mathrm{\tau}$ captures the suppression of the (net) diffusion due to clumping depth effects within the clump sublayer and $f_\mathrm{c}$ describes the likelihood that disc pebbles become "clumpy" when they reach the mid-plane. These factors will be further discussed in \Se{clumpmodel}. The other source term represents loss by clumps due to planetesimal formation
\begin{equation}\label{eq:sourcecp}
  S_\mathrm{plt} = \frac{\Sigma_\mathrm{c}}{t_\mathrm{plts}} = \epsilon\frac{\Sigma_\mathrm{c}}{t_\mathrm{sett}}
\end{equation}
where the settling time-scale is
\begin{equation}\label{eq:tsettle}
  t_\mathrm{sett} = \frac{1}{\mathrm{St} \Omega_K}
\end{equation}
and $\mathrm{St}$ is the Stokes number. In the clump pebble region, where the dust-to-gas ratio exceeds unity, planetesimal formation via gravitational collapse is expected \citep{JohansenEtal2006,ChiangYoudin2010,SimonEtal2016,SchoonenbergEtal2018}. We assume that the clumpy mid-plane dust forms planetesimals on a time-scale $t_\mathrm{plts} = t_\mathrm{sett}/\epsilon$ \citep{DrazkowskaEtal2016,SchoonenbergEtal2018,StammlerEtal2019}\hcadd{, where $t_\mathrm{sett}$ is the timescale on which filaments form via streaming instability in shearing box \citep{YangEtal2017}}. The planetesimal formation efficiency $\epsilon = 0.01$, unless otherwise stated, is taken similar to the value argued by \citet{DrazkowskaEtal2016} for the typical $\mathrm{St} \sim 0.01$ in our simulations. For smaller particles, the time-scales for settling and streaming instability increases\hcadd{\ \citep{YangEtal2017}}, while the formation rate of planetesimals is lower. After their formation, planetesimals do not interact with the other components.

\subsection{Disc Model: Gas and Disc Pebbles}\label{sec:discmodel}
 We consider gas and dust discs around $1\ {}M_\odot$ mass star. The gas surface density profile is set as 
\begin{equation}
    \Sigma_\mathrm{g} = \Sigma_\mathrm{g,0}\left(\frac{r}{100\,\mathrm{au}}\right)^{-1}.
\end{equation}
where $\Sigma_\mathrm{g,0}$ is taken $3\,\mathrm{g}\,\mathrm{cm}^{-2}$ in all of our simulations \hcadd{for simplicity, comparable with the gas surface density required to reproduce the DSHARP observations} \citep{ZhangEtal2018}. The gas disc in our simulations is taken isothermal in the vertical direction and does not evolve. The aspect ratio of the disc is chosen as
\begin{equation}\label{eq:cs}
    h =
  \frac{H_\mathrm{g}}{r} \equiv \frac{c_s}{\Omega_K r} = h_0\left(\frac{r}{100\,\mathrm{au}}\right)^{0.25} 
\end{equation}
where $c_s = \sqrt{k_B T/\mu m_p}$ is the isothermal sound speed with $m_p$ the proton mass and $\mu = 2.3$ the mean molecular weight in atomic units\hcrem{, and $\Omega_K(r) = \sqrt{GM_\star/r^3}$ is the local Keplerian frequency}. The aspect ratio at 100 au, $h_0$, is taken $0.075$ unless otherwise stated. \Eq{cs} corresponds to a disc temperature profile of
\begin{equation}\label{eq:Td}
    T = \frac{\mu m_p GM_\star h^2}{k_B r} = T_0\left(\frac{r}{100\,\mathrm{au}}\right)^{-0.5}
\end{equation}
where $T_0 = 13.9$ K is the disc temperature at $100$ au for a solar mass star.

At the outer boundary disc pebbles flow in at imposed rates $\dot{M}_\mathrm{ext}$. The initial disc pebble surface density is taken to be the steady-state value
\begin{equation}\label{eq:Sigp0}
  \Sigma_\mathrm{d} = \frac{\dot{M}_\mathrm{ext}}{2\pi r v_\mathrm{dr}}
\end{equation}
The external mass flux $\dot{M}_\mathrm{ext}$ is a key model parameter.

In the Epstein drag regime, -- particle radius less than mean free path; a condition valid for the disc and dust parameters of interest here -- the normalized stopping time of particles with radius $a_\bullet$ and internal density $\rho_\bullet$ in the mid-plane of the disc is the Stokes number \citep{Weidenschilling1977,BirnstielEtal2012}
\begin{equation}\label{eq:St}
    \mathrm{St} = \frac{\pi}{2}\frac{\rho_\bullet a_\bullet}{\Sigma
    _\mathrm{g}}
\end{equation}

We consider two setups:
\begin{enumerate}
    \item runs characterized by a constant Stokes number (the particle size changes with orbital radius); 
    \item runs characterized by constant particle size (the Stokes number changes with radius). 
\end{enumerate}
The former is consistent with a scenario in which particles attain their size by a velocity fragmentation threshold \citep{BirnstielEtal2010}; the latter by a bouncing threshold \citep{GuettlerEtal2010}. We summarize these parameters in \Tb{models}.

Using the $\alpha$-prescription for vertical transport of the disc pebbles with dimensionless \hcch{viscosity}{diffusivity} parameter $\delta_\mathrm{d}$, we set $\delta_\mathrm{d} = 10^{-3}$ in all of our simulations \hcadd{as suggested from the vertical extent of dust in edge-on observations of ALMA disks \citep[e.g.,][]{PinteEtal2016,VillenaveEtal2020}. The disk pebbles' vertical diffusivity can be different from their radial diffusivity defined earlier (see \Se{assessment} for discussion)}. Assuming a vertical Gaussian distribution for both disc pebble and gas density, the pebble disc scale height
\begin{equation}\label{eq:Hd}
    H_\mathrm{d} = \sqrt{\frac{\delta_\mathrm{d}}{\delta_\mathrm{d}+\mathrm{St}}}H_\mathrm{g}
\end{equation}
is obtained by balancing dust settling with vertical diffusion \citep{DubrulleEtal1995,BirnstielEtal2010}.

\subsection{Clump Model}\label{sec:clumpmodel}
Following \citet{Sekiya1998} and \citet{YoudinShu2002}, we assume that clump pebbles reside in the mid-plane region with the scale height
\begin{equation}\label{eq:Hc}
  H_\mathrm{c} = \sqrt{Ri_\mathrm{crit}} \eta r \Psi(\psi)
\end{equation}
hereafter referred to as the clump layer. For the critical Richardson number $Ri_\mathrm{crit}$, \citet{Chiang2008} points out that when the vertically-integrated dust-to-gas ratio is between 1 and 5 times solar abundance, the Richardson number $Ri \sim 0.1$. Thus we take $Ri_\mathrm{crit} = 0.1$. The dimensionless self-gravitational term $\Psi(\psi) \equiv \sqrt{1+2\psi} - \psi \ln{[(1+\psi+\sqrt{1+2\psi})/\psi]}$ is an order of unity, where $\psi \equiv 4\pi G \rho_\mathrm{g}/\Omega_K^2 = 0.23 (r/100\,\mathrm{au})^{0.75}$ is the self-gravity of gas. By equating the diffusion time-scale from clump pebbles to disc pebbles
\begin{equation}
  t_\mathrm{diff} = \frac{H_\mathrm{c}^2}{\alpha_\mathrm{c}c_s H_\mathrm{g}}
\end{equation}
and the settling time-scale (\Eq{tsettle}), we can calculate the sublayer diffusivity with default values of parameters for clumps. Accordingly, we choose $\alpha_\mathrm{c} = 10^{-5}$ in the thin mid-plane layer in our default model (see \se{assessment} for a more detailed discussion).

Spontaneous local overdense clumping of dust by streaming instability was firstly found by \citet{JohansenEtal2006} and \citet{JohansenYoudin2007}\hcrem{, where it shows vertically stratified dust layer}. A transition from a homogeneous to a clumpy medium happens at volume dust-to-gas ratio equal to unity \citep{JohansenYoudin2007}. And strong radial concentration of solids suddenly operates, leading to roughly axisymmetric substructures \hcadd{with vertically stratified dust layer} \citep{YangJohansen2014,LiEtal2018}. To qualitatively mimic the appearance of clumps, we introduce the efficiency factor
\begin{equation}\label{eq:fsek}
  f_\mathrm{c} = 1 - \exp{(-Z_\mathrm{mid}^3)}
\end{equation}
where the mid-plane dust-to-gas ratio is
\begin{equation}\label{eq:Zmid}
    Z_\mathrm{mid} = \frac{\rho_\mathrm{d,0}}{\rho_\mathrm{g,0}} = Z_\mathrm{mid,dd}+Z_\mathrm{mid,cd} =  \frac{\Sigma_\mathrm{c}}{\Sigma_\mathrm{g}}\frac{H_\mathrm{g}}{H_\mathrm{c}} + \frac{\Sigma_\mathrm{d}}{\Sigma_\mathrm{g}}\frac{H_\mathrm{g}}{H_\mathrm{d}}
\end{equation}
in which $\rho_\mathrm{d,0}$ and $\rho_\mathrm{g,0}$ are the mid-plane density of dust and gas respectively. When the dust-to-gas ratio is $\ll$1, \eq{fsek} simply implies that a clumpy mid-plane layer will not form: $f_\mathrm{c}\ll1$ and there is no disc-to-clump conversion. Otherwise, when $Z_\mathrm{mid}\gtrsim1$, a clump layer is present and disc pebbles transform into clump pebbles on a settling time-scale. In order for the ring to survive, the transition of $f_\mathrm{c}$ from 0 to 1 must be rapid; otherwise the dichotomy between a clumpy ring and free-drifting dust will not be established.  Exponents larger than $2$ in \eq{fsek} will not change the conclusions of this work significantly.

Apart from planetesimal formation, vertical diffusive transport from the clump layer to the disc pebble component also contributes to the loss of clump pebble. However, the rate at which the clump sublayer "erodes" depends on the drain of clumping in clump layer. Pebbles, liberated from dispersing clumps, may yet be prevented from escaping the clump layer, as they can collide with other particles in the clump sublayer \citep{KrijtCiesla2016}. 

This suppression of diffusion by the collisional "self-shielding" is modeled through a reduction in the diffusivity by a factor $f_\tau$ \eq{ftau}:
\begin{equation}
  f_\tau = \frac{1-e^{-\tau}}{\tau}.
\end{equation}
where $\tau$ is the "optical depth" of clump layer. In the optical thin limit ($\tau\ll1$) $f_\tau \simeq 1$ and there is no reduction: pebbles in the clump layer can freely exchange with disc pebbles through diffusion. In the other limit ($\tau\gg1$) $f_\tau \simeq 1/\tau$, meaning that only a fraction of the clump layer ($\Sigma_\mathrm{c}/\tau$) is able to diffuse out of the clump mid-plane without being impeded. That is, only the surface layers do exchange material. This is analogous to the escape of radiation from layers of optical depth concept below unity.

For the clump depth of the mid-plane layer, one would naively write $\tau = \kappa_\bullet \Sigma_\mathrm{c}$ where $\kappa_\bullet$ the "collision opacity" of the pebbles
\begin{equation}\label{eq:kappadot}
    \kappa_\bullet \equiv \frac{4\pi a_\bullet^2}{m_\bullet} = \frac{3}{\rho_\bullet a_\bullet}.
\end{equation}
A complication, however, is the clumpiness of the medium as the medium becomes more transparent when particles clump. To encapsulate these effect, we need to present a macroscopic clump model. In Appendix \ref{app:clumps} we present a simple clump model, accounting for these effects. In this model, it is assumed that the physical properties of clumps (their size, filling factor, etc) are constant, such that their density increases with mass loading ($Z_\mathrm{mid}$ or $\Sigma_\mathrm{c}$). Then, at a critical loading the clumps saturate and no longer contribute to the effective clumping depth of the medium. Correspondingly, we write for the effective clumping depth
\begin{equation}
    \tau_\mathrm{eff} = \min (\kappa_\bullet \Sigma_\mathrm{c}, \tau_\mathrm{max})
\end{equation}
and use this expression in \eq{ftau} to evaluate the reduction in diffusivity factor $f_\tau$. The parameter $\tau_\mathrm{max}$ signifies at which point clumping starts to make the medium more transparent. When $\tau_\mathrm{max}<1$ it means that the clumps fill a small fraction of the space and that the particles liberated from dispersing clumps would easily escape, which does not favour ring survival. On the other hand $\tau_\mathrm{max}\gg1$ means that collisional reduction of the diffusion \citep{KrijtCiesla2016} would promote the survival of clumping state, which is the desired situation. In this paper we adopt $\tau_\mathrm{max}=15$. Although an (unknown) free parameter, we have verified that the conclusions of this paper are unaffected as long as $\tau_\mathrm{max}\gg1$.

\begin{table*}
\caption{Model runs.\label{tab:models} (1) Stokes number; (2) particle size; (3) initial clump perturbation contrast to background disc pebble density; (4) advection velocity for clumps; (5) external mass flux; (6) planetesimal formation efficiency; (7) clump pebble turbulence $\alpha$-parameter; (8) aspect ratio at $100$ au (9) result of rings; (10) predicted velocity of the ring; (11) measured velocity of the ring; (12) predicted leaking mass flux; (13) measured leaking mass flux; \hcadd{(14) section reference where the model is mainly discussed.}}
\centering
\small
\begin{tabular}{l|cccccccc|lccccc}
\hline\hline
model-id & $\mathrm{St}$ & $a_\bullet$ & $A$ & $f_\mathrm{ca}$$^a$ & $\dot{M}_\mathrm{ext}$ & $\epsilon$ & $\alpha_\mathrm{c}$ & $h_0$ & outcome & $v_\mathrm{ring}$ & $v_\mathrm{ring,ms}$ & $\dot{M}_\mathrm{leak}$ & $\dot{M}_\mathrm{leak,ms}$ &\\
& & [$\mu$m] & & & [$\mathrm{M}_\oplus\,\mathrm{Myr}^{-1}$] & & & & & \multicolumn{2}{c}{[$\,\mathrm{au}\,\mathrm{Myr}^{-1}$] } & \multicolumn{2}{c}{[$M_\oplus,\mathrm{Myr}^{-1}$]} &\\
& (1) & (2) & (3) & (4) & (5) & (6) & (7) & (8) & (9) & (10) & (11) & (12) & (13) & (14)\\
\hline
\texttt{t1f1} & $0.01$$^b$ &   & 1 & 1 & 100 & $10^{-2}$ & $1\times10^{-5}$ & 0.075 & ring & 10 & 1.3 & 40 & 65 & \ref{sec:cnst-Stokes}\\
\texttt{t1f1A8} & $0.01$ &   & 0.8 & 1 & 100 & $10^{-2}$ & $1\times10^{-5}$ & 0.075 & ring & 10 & 1.3 & 40 & 65 & \ref{sec:cnst-Stokes}\\
\texttt{t1f1A6} & $0.01$ &   & 0.6 & 1 & 100 & $10^{-2}$ & $1\times10^{-5}$ & 0.075 &   & 10 &   & 40 &  & \ref{sec:cnst-Stokes}\\
\texttt{t1f1c9} & $0.01$ &   & 1 & 0.9$^c$ & 100 & $10^{-2}$ & $1\times10^{-5}$ & 0.075 & ring & 10 & -7.2 & 40 & 65 & \ref{sec:ringmig}\\
\texttt{t1f1c6} & $0.01$ &   & 1 & 0.6 & 100 & $10^{-2}$ & $1\times10^{-5}$ & 0.075 & ring & 10 & -35.5 & 40 & 75 & \ref{sec:ringmig}\\
\texttt{t1f1c3} & $0.01$ &   & 1 & 0.3 & 100 & $10^{-2}$ & $1\times10^{-5}$ & 0.075 &   & 10 &   & 40 &  & \ref{sec:ringmig}\\
\texttt{t1f1c0} & $0.01$ &   & 1 & 0 & 100 & $10^{-2}$ & $1\times10^{-5}$ & 0.075 &   & 10 &   & 40 &  & \ref{sec:ringmig}\\
\texttt{t1f2} & $0.01$ &   & 1 &   & 100 & $10^{-2}$ & $1\times10^{-5}$ & 0.075 & ring & 10 & 9.5 & 40 & 30 & \ref{sec:ringmig}\\
\texttt{t1f1h6} & $0.01$ &   & 1 & 1 & 100 & $10^{-2}$ & $1\times10^{-5}$ & 0.06 & ring & 12 & 6.7 & 20 & 36 & \ref{sec:ringmig}\\
\texttt{t1f1h8} & $0.01$ &   & 1 & 1 & 100 & $10^{-2}$ & $1\times10^{-5}$ & 0.08 & ring & 9 & $-1.0$ & 49 & 78 & \ref{sec:ringmig}\\
\texttt{t1f1h9} & $0.01$ &   & 1 & 1 & 100 & $10^{-2}$ & $1\times10^{-5}$ & 0.09 &   & 8 &   & 69 &  & \ref{sec:ringmig}\\
\texttt{t1f2h6} & $0.01$ &   & 1 &   & 100 & $10^{-2}$ & $1\times10^{-5}$ & 0.06 & ring & 12 & 13.5 & 20 & 18 & \ref{sec:ringmig}\\
\texttt{t1f2h8} & $0.01$ &   & 1 &   & 100 & $10^{-2}$ & $1\times10^{-5}$ & 0.08 & ring & 9 & 8.4 & 49 & 35 & \ref{sec:ringmig}\\
\texttt{t1f2h9} & $0.01$ &   & 1 &   & 100 & $10^{-2}$ & $1\times10^{-5}$ & 0.09 & ring & 8 & 6.2 & 69 & 46 & \ref{sec:ringmig}\\
\texttt{t1f2a3} & $0.01$ &   & 1 &   & 100 & $10^{-2}$ & $1\times10^{-3}$ & 0.075 &   & 100 &   & 40 &  & \ref{sec:ringmorph}\\
\texttt{t1f2a4} & $0.01$ &   & 1 &   & 100 & $10^{-2}$ & $1\times10^{-4}$ & 0.075 & ring & 32 & 24.4 & 40 & 50 & \ref{sec:ringmorph}\\
\texttt{t1f2a6} & $0.01$ &   & 1 &   & 100 & $10^{-2}$ & $1\times10^{-6}$ & 0.075 & ring & 3 & 3.2 & 40 & 30 & \ref{sec:ringmorph}\\
\texttt{t1f1e1} & $0.01$ &   & 1 & 1 & 100 & $10^{-1}$ & $1\times10^{-5}$ & 0.075 &   & 10 &   & 40 &  & \ref{sec:ringsuv}\\
\texttt{t1f1e3} & $0.01$ &   & 1 & 1 & 100 & $10^{-3}$ & $1\times10^{-5}$ & 0.075 & ring & 10 & 2.1 & 40 & 65 & \ref{sec:ringsuv}\\
\texttt{t1f1e4} & $0.01$ &   & 1 & 1 & 100 & $10^{-4}$ & $1\times10^{-5}$ & 0.075 & ring & 10 & 2.2 & 40 & 65 & \ref{sec:ringsuv}\\
\texttt{t1f2e1} & $0.01$ &   & 1 &   & 100 & $10^{-1}$ & $1\times10^{-5}$ & 0.075 & ring & 10 & 6.1 & 40 & 30 & \ref{sec:ringsuv}\\
\texttt{t1f2e3} & $0.01$ &   & 1 &   & 100 & $10^{-3}$ & $1\times10^{-5}$ & 0.075 & ring & 10 & 9.7 & 40 & 30 & \ref{sec:ringsuv}\\
\texttt{t1f2e4} & $0.01$ &   & 1 &   & 100 & $10^{-4}$ & $1\times10^{-5}$ & 0.075 & ring & 10 & 9.7 & 40 & 30 & \ref{sec:ringsuv}\\
\texttt{t1f3} & $0.01$ &   & 1 & 1 & 50 & $10^{-2}$ & $1\times10^{-5}$ & 0.075 &   & 5 &   & 40 &   & \ref{sec:ringsuv}\\
\texttt{t1f4} & $0.01$ &   & 1 & 1 & 200 & $10^{-2}$ & $1\times10^{-5}$ & 0.075 & ring & 20 & 8.2 & 40 & 65 & \ref{sec:ringsuv}\\
\texttt{t2f1} & $0.02$ &   & 1 & 1 & 100 & $10^{-2}$ & $1\times10^{-5}$ & 0.075 &   & 5 &   & 80 &   & \ref{sec:ringsuv}\\
\texttt{t3f1} & $0.005$ &   & 1 & 1 & 100 & $10^{-2}$ & $1\times10^{-5}$ & 0.075 & ring & 28 & 6.3 & 20 & 30 & \ref{sec:ringsuv}\\
\texttt{s1f1} &   & 103$^d$ & 1 & 1 & 100 & $10^{-2}$ & $1\times10^{-5}$ & 0.075 & ring &   &   &   &   & \ref{sec:cnst-size}\\
\texttt{s1f1A0} &   & 103 & 0 & 1 & 100 & $10^{-2}$ & $1\times10^{-5}$ & 0.075 & ring &   &   &   &   & \ref{sec:cnst-size}\\
\texttt{s2f2} &   & 1030 & 1 & 1 & 400 & $10^{-2}$ & $1\times10^{-4}$ & 0.075 & Cycle &   &   &   &   & \ref{sec:cnst-size}\\
\texttt{Elias24} &   & 93.4 & 0.5 &   & 200 & $10^{-2}$ & $3\times10^{-5}$ & 0.075 & Elias 24 &   &   &   &  & \ref{sec:E24}\\
\texttt{AS209} &   & 180 & 0.2, 1$^e$ &   & 300 & $10^{-2}$ & $1\times10^{-5}$ & 0.075 & AS 209 &   &   &   &   & \ref{sec:AS209}\\
\hline
\multicolumn{14}{l}{\footnotesize$^a$ "1" for runs where clump advection follows the NSH solution. Empty for runs where there is no clump advection, \hcadd{see \se{transportmodel}.}}\\
\multicolumn{14}{l}{\footnotesize$^b$ constant Stokes number, particle size changes with location}\\
\multicolumn{14}{l}{\footnotesize$^c$ fraction of NSH velocity in clump drift, see \eq{fca}}\\
\multicolumn{14}{l}{\footnotesize$^d$ constant particle size, Stokes number is variable. $103\,\mu$m corresponds to $\mathrm{St} = 0.01$ at $r = 100$ au}\\
\multicolumn{14}{l}{\footnotesize$^e$ two initial perturbation are set in this sample, see \se{AS209}}
\end{tabular}
\end{table*}

\section{Results}\label{sec:results}
We conduct two classes of simulations, models with constant Stokes number and models with constant particle sizes. The parameters are summarized in \Tb{models}. \hcadd{Column (1) to (8) are input parameters of the model. Column (9) to (13) are output and corresponding measured values. We list the section in which each model is mainly discussed in Column (14).}  The grid resolution is fixed at $\Delta x = 0.005$ au. For our default model, the outer boundary is at $150$ au and the inner boundary is at $50$ au. The gas density profile does not evolve in time. The evolution for each dust component is determined by the transport equations described above. The dust outer boundary is set according to a fixed external mass flux $\dot{M}_\mathrm{ext}$. At the inner boundary of the domain, a free-outflow condition is set. We first describe our default model with constant Stokes number in \se{cnst-Stokes}\hcadd{. We quantify the mass flux leaking from the ring downstream in \se{Mleak}} and discuss the motion of the ring in \se{ringmig}\hcadd{ and the morphology of the ring in \se{ringmorph}}. Then we conduct a parameter study about the survival of the ring in \se{ringsuv}. In \se{cnst-size}, we turn to the constant size model.

\begin{figure}
    \centering
    \includegraphics[width=0.99\columnwidth]{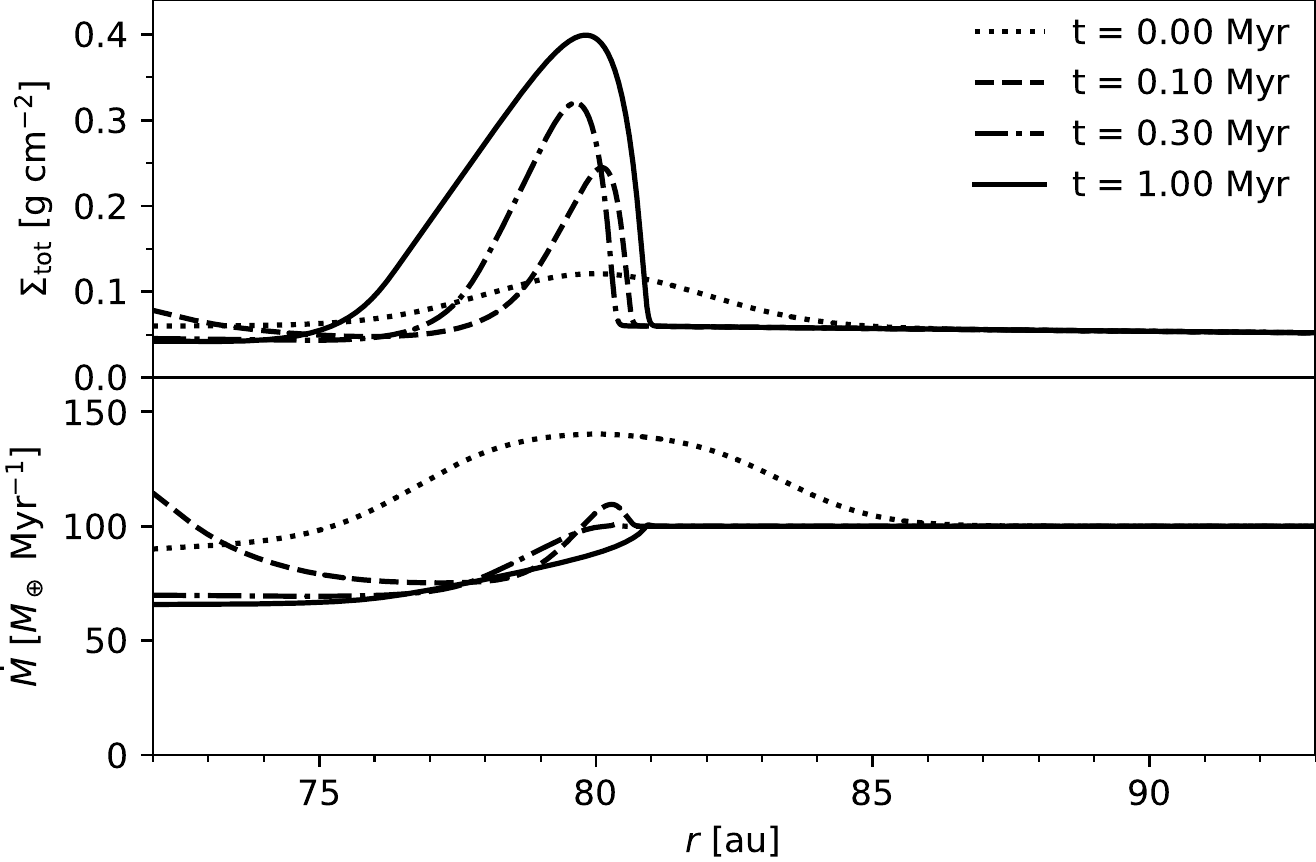}
    \includegraphics[width=0.99\columnwidth]{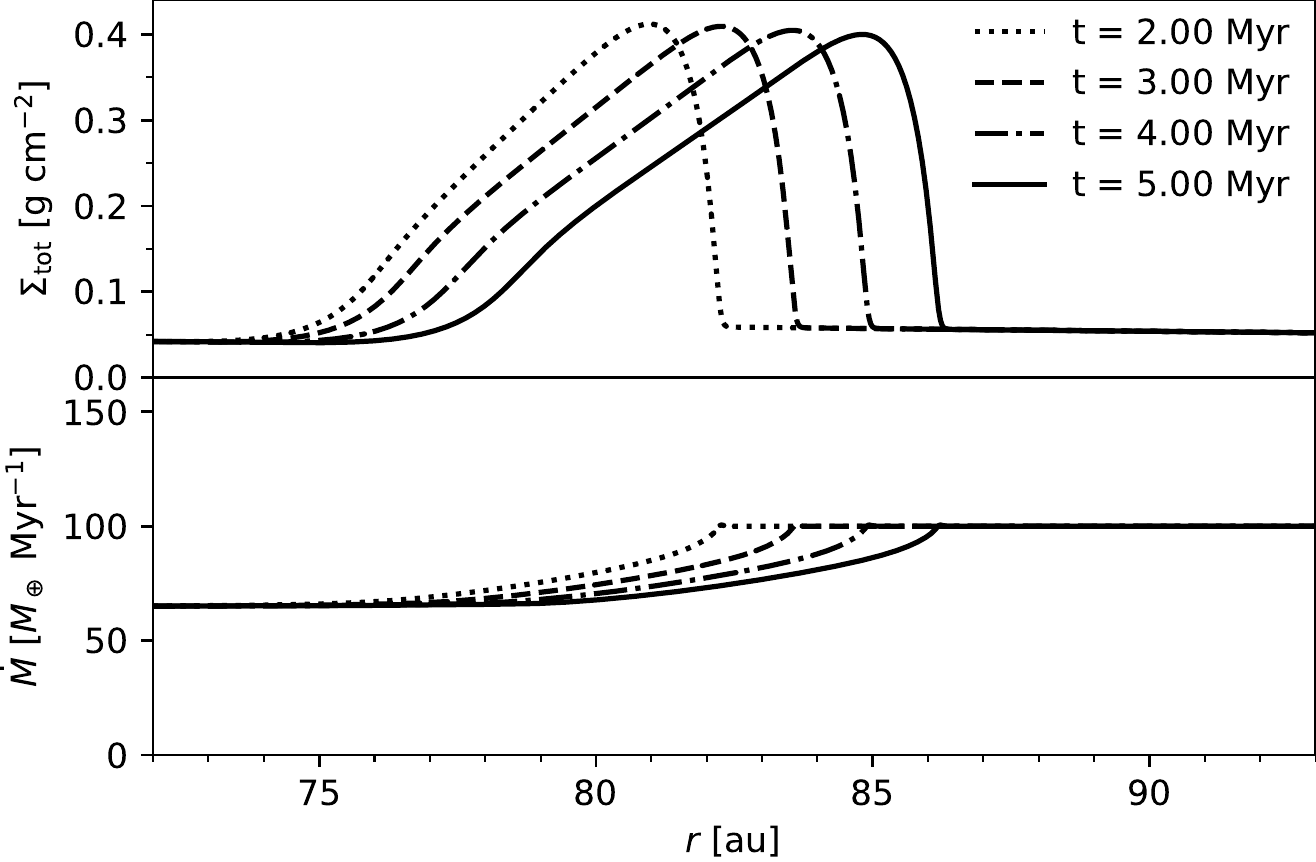}
    \caption{\label{fig:t1f1-evo} Dust radial profile and mass flux of default model \texttt{t1f1} for the ring growth phase (upper two panels) and later evolution phase (lower two panels). The Stokes number is constant at $\mathrm{St} = 0.01$. Ring growth happens within $\approx$1 Myr. After that, the ring's shape stays self-similar, while moving outward at a velocity of $\approx$1$\,\mathrm{au\,Myr}^{-1}$. Other parameters are listed in \Tb{models}}.
\end{figure}

\ccc{DISCUSSION ON MODEL PARAMETERS}

\subsection{Constant Stokes Runs}\label{sec:cnst-Stokes}
In our default model \texttt{t1f1}, we fix the Stokes number at 0.01, and set the constant external mass flux $100\ \mathrm{M}_\oplus\,\mathrm{Myr}^{-1}$ \hcadd{(See \se{assessment} for assessment)}. The planetesimal formation efficiency is $\epsilon = 0.01$ such that $1\%$ of the clump collapse into planetesimal in a settling time-scale. Dimensionless diffusivities are $\delta_\mathrm{d} = 10^{-3}$ for the disc pebbles and $\alpha_\mathrm{c} = 10^{-5}$ for the clumps respectively. At $t=0$, the simulation is initialized by inserting a Gaussian-shaped "ring" of clump material on top of the smooth power-law dust disc. The disc pebble background plus the radial Gaussian distribution of the clump give the total dust density:
\begin{equation}\label{eq:sigmatot}
  \Sigma_\mathrm{tot}(r) = \Sigma_\mathrm{d} + \Sigma_\mathrm{c} = \Sigma_\mathrm{d}(r) + A\,
  \Sigma_0 
  \exp\left(-\dfrac{1}{2}\frac{(r-r_0)^2}{w^2}\right)
\end{equation}
where the reference radius $r_0 = 80$ au, and $\Sigma_0 = \Sigma_\mathrm{d}(r_0)$ (\eq{Sigp0}). Hence, the relative amplitude $A = 1$ means that the peak of total dust density is twice that of the background value. The width of the initial perturbation is taken as $w = 2$\,au. The main parameters are listed in \Tb{models}.

 We ran the simulation for $5$ Myr. We divide the evolution of this ring into two stages: the ring growth phase and the later evolution phase. The upper two panels of \fg{t1f1-evo} show the radial profile of the total dust surface density and the mass flux during the ring growth phase. Within $\approx$0.1 Myr, the initial perturbation transforms into a ring-like shape. Since the differential radial drift caused by concentration of clump around the initial perturbation location of $r =80$ au, a pileup of pebbles happens and make the ring narrow. During $\sim$0.1--$1$ Myr, disc pebbles are being "eaten" by the ring, and ring grows as clump pebbles accumulate locally. The height and width of the ring are still evolving but the location of the ring peak does not change significantly. As the concentration of dust increases inside the ring, the measured mass flux leaking from the ring downstream gradually decreases to approach $65\ M_\oplus$ Myr$^{-1}$, which is lower than the external mass flux because of consumption of the ring and planetesimal formation inside. At $\sim 1$ Myr, the ring shape has reached equilibrium reflecting a balance between the consumption of disc pebbles by the ring and the loss due to planetesimal formation.
 
  \begin{figure}
    \centering
    \includegraphics[width=0.99\columnwidth]{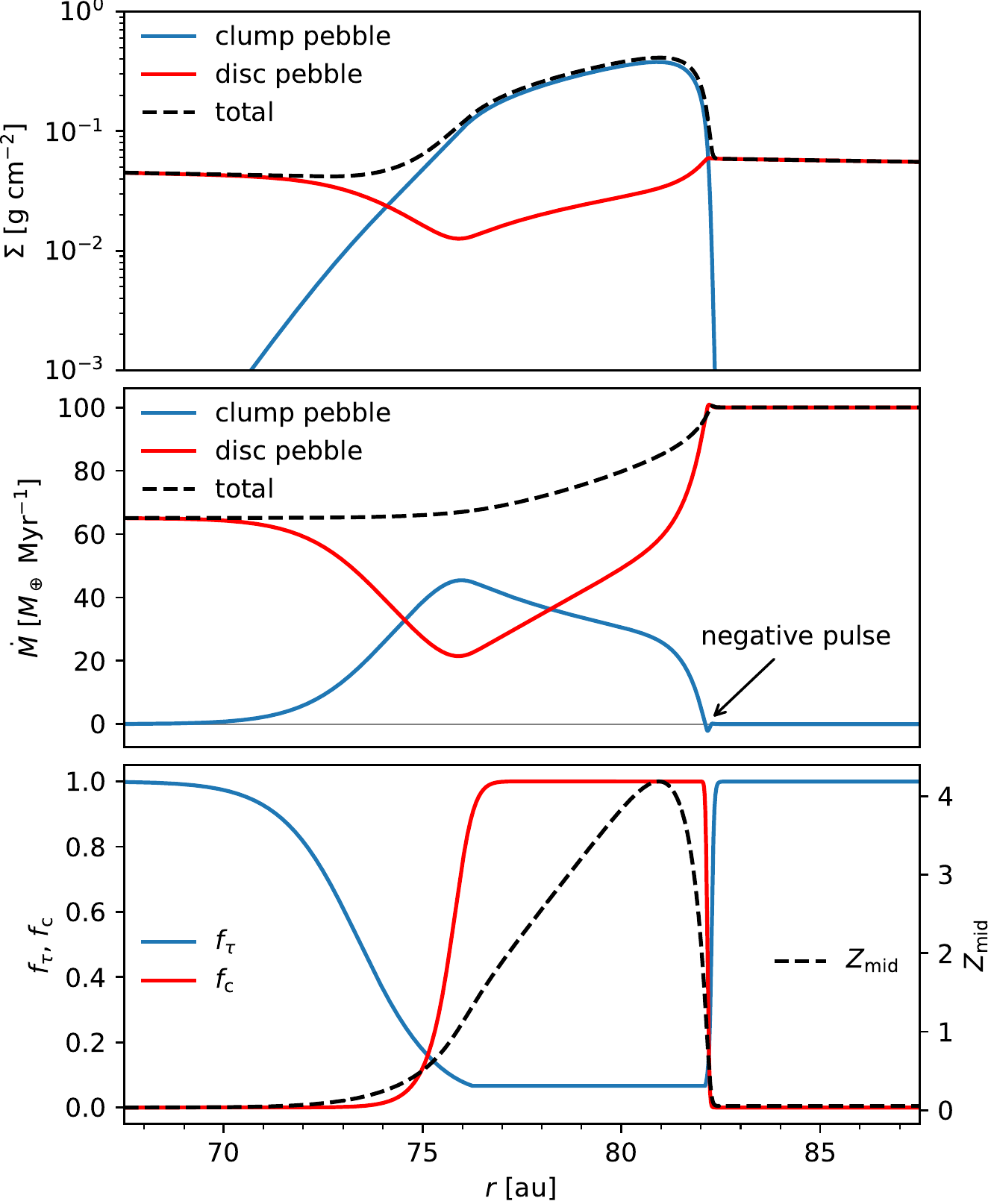}
    \caption{\label{fig:t1f1-factor}Snapshots of run \texttt{t1f1} at $t=2$ Myr,\protect\\ \textbf{Up:} disc pebble (red) and clump (blue) surface density distribution around the ring. \textbf{Middle:} mass flux contributed by the two components. Clumps exist only inside the ring and keep moving inward. The outward motion of the ring is not an overall migration of the dust, but a phase velocity effect. The solid black arrow marks the point where negative mass flux of clump at ring's outer boundary is observed due to the steep $\Sigma$-gradient. \textbf{Down:} diffusion suppression factor $f_\tau$ (blue), settling factor $f_\mathrm{c}$ (red) and mid-plane dust-to-gas ratio $Z_\mathrm{mid}$(black). Low $f_\mathrm{c}$ indicates suppression from disc to clump pebble conversion and low $f_\tau$ represent inefficient transport of the clump sublayer.}
\end{figure}

A sufficiently strong seed ring is required to get sustained clumping. Otherwise the initial perturbation will just fade away and all the dust will end up in the disc state. In order to assess this quantitatively, we decrease the initial amplitude $A$ and find that the bifurcation point lies around $A=0.7$ for the parameter setup of \texttt{t1f1}. Once the initial perturbation's amplitude exceeds this value, further increasing $A$ will not affect the ring characteristics. \hcadd{(see \texttt{t1f1A8}, \texttt{t1f1A6} in \Tb{models})}

To better understand the material exchange, we show in \Fg{t1f1-factor} the surface density and radial mass flux contributed by disc pebbles and clump pebbles, respectively, at $t = 2$ Myr. From the upper panel, it is clear that clumps only exist in the ring, where they constitute the vast majority of the ring's mass. The high concentration of clumps within the ring region slows down their drift according to \eq{vdr}; and the faster drifting disc pebbles are "eaten" by the slowly drifting ring. Therefore, both the disc pebble density and pebble flux are lower in the ring than outside the ring. \Fg{t1f1-factor} also shows the diffusion suppression factor $f_\tau$ and the settling factor $f_\mathrm{c}$, which are indicators for the presence of the ring. The settling factor $f_\mathrm{c}$ quantifies the ability of disc dust to transfer into clumpy mid-plane dust, while the diffusion suppression factor $f_\tau$ quantifies the ability of clump material to diffuse back to the disk component. Both of these factors depend on the mid-plane dust-to-gas ratio (\eq{Zmid}; dashed line). Because $Z_\mathrm{mid}\gg1$ withing the ring region, $f_\mathrm{c}\approx1$ over a broad spatial scale, allowing the incoming disk dust to be incorporated into the clumpy mid-plane. In addition, $f_\tau$ has reached its minimum within the disk region, signifying that the clumpy mid-plane has become "optical thick" to collisions and that exchange to the disk component only happens at the surface, since the liberated pebbles bump into each other and cannot escape the clump layer (see \se{clumpmodel}).

As a result of these "clumpy effects", pebbles in the disc turn into nearly-stationary clumpy state efficiently and are trapped in clump layer. After the ring's growing phase, the combined surface density of pebble and clump dust stops increasing and the morphology of the ring stabilizes. The ring starts to "migrate" outward at speed $\sim $ au$/$Myr as shown in the lower two panels of \fg{t1f1-evo}. The reason for the outward movement is that clump material is being added to the front side of the ring and lost at the back, whereas the ring as a whole suffers little from inward drifting due to its high concentration. At the outer edge of the ring, diffusion of clump pebbles catches up with the upstream disc pebble flow, whereas flux leaks from the ring downstream. As a result, the ring's morphology stays preserved and we see the ring "moving" outward. A more detailed discussion of the ring motion is given in \se{ringmig}.
 
\begin{figure}
    \centering
    \includegraphics[width=0.99\columnwidth]{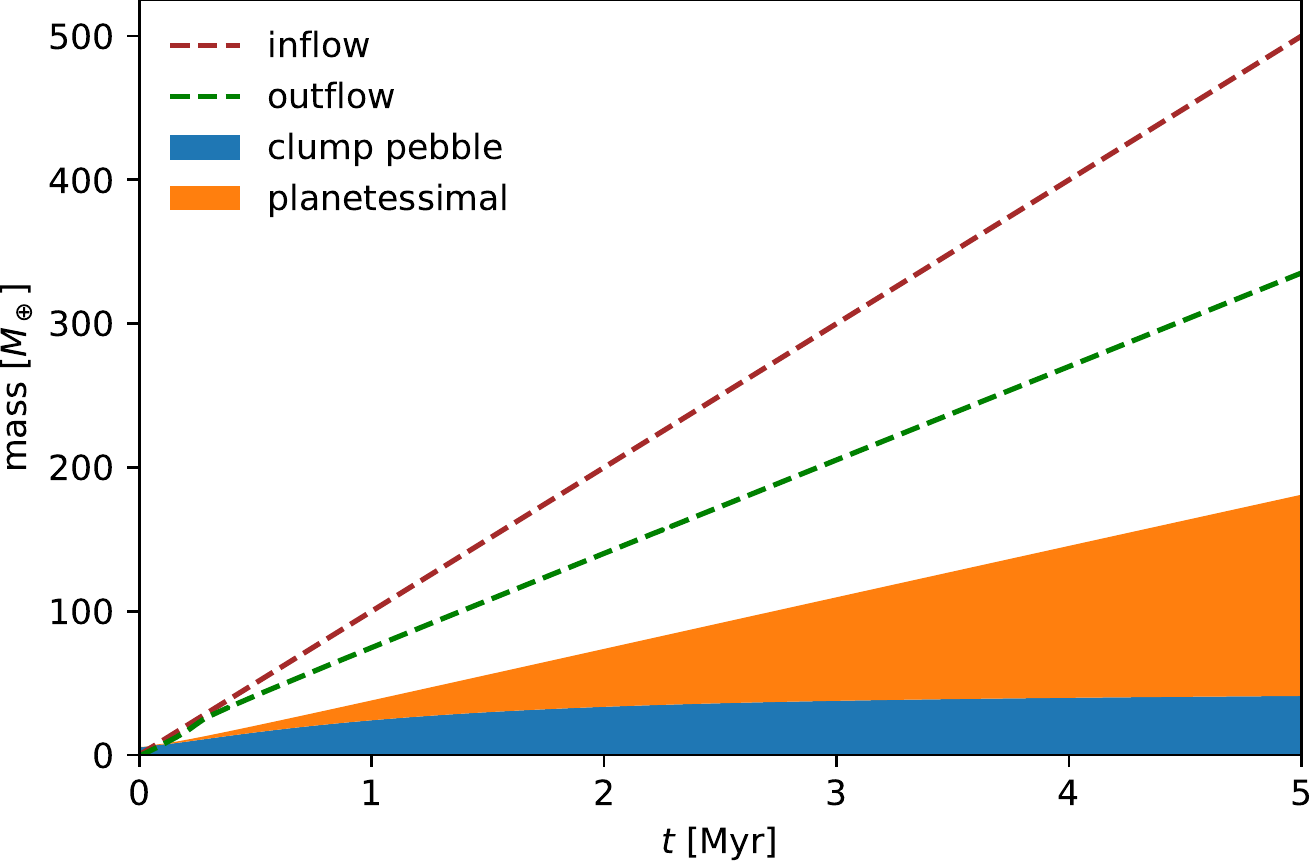}
    \caption{\label{fig:t1f1-mass}Time evolution of the mass budget. The red and green line show the accumulated inflow and outflow of mass respectively. The slope of the line denoted "inflow" corresponds to the external mass flux. The blue block represents the mass of clump pebbles in the simulation at time $t$ and the orange block on top represents the total mass in planetesimals. After $t=2$ Myr the ring's mass (clump pebble mass) has converged to $\simeq\!35\,M_\oplus$.}
\end{figure}

We show the mass budget evolution of different dust components in \fg{t1f1-mass}. The red and green line show the accumulated inflow and outflow of mass respectively. The blue region is the mass of clump pebbles in the simulation at time $t$, which correspond to the mass of ring. The orange region shows the accumulated mass of formed planetesimal. The inflow slope is constant at $100M_\oplus\ \mathrm{Myr}^{-1}$ consistent with the external mass flux. Starting with a small amount of mass, the clump pebble mass first increase rapidly during the ring growth phase. In the mean time, the outflow mass and planetesimal mass is still evolving as well. At $t=2\,$Myr, both the outflow and planetesimal formation rate converge to constant values. Together with the constant external mass flux, the mass budget of different components enters equilibrium -- the mass of the ring (clump mass) saturates to $\simeq\!35\,M_\oplus$ and the planetesimal production approaches $\simeq\!30\,M_\oplus/$Myr. After $5$ Myr, the amount of pebbles inside the ring is $41\ M_\oplus$ and the planetesimal belt is $140\ M_\oplus$ massive. As shown by the plot, the clumpy ring does not absorb all the incoming mass flux but continue to leak a significant mass to the inner disk.

\subsection{The Leaking Mass Flux}\label{sec:Mleak}
Here, we use a simple model to quantify the value of the leaking mass flux $\dot{M}_\mathrm{leak}$ -- the rate at which the ring loses material to the downstream regions.

We consider the inner edge of the ring at the boundary between the clumpy ring and freely-drifting downstream. We assume that near this point, $Z\approx Z_\mathrm{crit} =1$, most of the clump material gets released. The corresponding clump density is
\begin{equation}\label{eq:balance}
    \frac{\rho_\mathrm{c}}{\rho_\mathrm{g}} = \frac{\Sigma_\mathrm{c}}{\Sigma_\mathrm{g}}\frac{H_\mathrm{g}}{H_\mathrm{c}} = Z_\mathrm{crit} = 1.
\end{equation}
For simplicity, the mid-plane contributed by disc pebbles is neglected comparing to the clumps'. Further assuming that all of the clumps at the this point will be converted into radially-drifting disc pebbles, the surface densities of disc pebble should equal the released clump pebbles. Defining the leaking mass flux as $\dot{M}_\mathrm{leak} = 2\pi r v_\mathrm{dr,d} \Sigma_\mathrm{d}$ we obtain using the expressions in \eq{balance}
\begin{equation}\label{eq:M_leak}
    \dot{M}_\mathrm{leak} = 2\pi r v_\mathrm{dr,d} \Sigma_\mathrm{g} \frac{H_\mathrm{c}}{H_\mathrm{g}}
\end{equation}
Since the time-scale of materials exchange is comparable to the drift time-scale pass the the ring, the exact location of the edge is uncertain in our simulations. We take a numerical factor $f_\mathrm{edg}$ to quantify the uncertainty introduced by choosing $Z=1$ for the edge. With further substitutions for $v_\mathrm{dr,d}$, $H_\mathrm{c}$ and $H_\mathrm{g}$, we obtain
\begin{equation}\label{eq:Mleak}
\begin{aligned}
    \dot{M}_\mathrm{leak} &\simeq f_\mathrm{edg} \mathrm{St} \Sigma_\mathrm{g,0} h_0^3 M_\star^{0.5} r^{0.25} \\ 
    & = 40\,M_\oplus\,\mathrm{Myr}^{-1} f_\mathrm{edg} \left(\frac{\mathrm{St}}{10^{-2}}\right) \left(\frac{\Sigma_\mathrm{g,0}}{3\,\mathrm{g}\,\mathrm{cm}^{-2}}\right) \\
    & \times\left(\frac{h_0}{0.075}\right)^3 \left(\frac{M_\star}{M_\odot}\right)^{0.5} \left(\frac{r}{80\,\mathrm{au}}\right)^{0.25}
\end{aligned}
\end{equation}
In \Tb{models}, we compute the leaking mass flux predicted from \eq{Mleak}, and measure the exact mass flux downstream for each run as $\dot{M}_\mathrm{leak,ms}$. By comparison \eq{Mleak} to the measured $\dot{M}_\mathrm{leak,ms}$ we obtain that $f_\mathrm{edg}$ is within $[0.6,0.8]$ for runs where clumps have no advection term and within $[1.5,1.6]$ for runs with clump drift (see more comperision in \se{ringmig}). The dependence on Stokes number and gas surface density signify that larger particles are easier to leak from the ring. The expression shows that the leaking mass flux is rather sensitive to the aspect ratio (temperature) of the disc. Thicker/hotter discs result in less efficient pileups because smaller mid-plane concentration and faster pebble drifting. In order for rings to survive, the mass flow entering the ring region must exceed $\dot{M}_\mathrm{leak}$. Therefore, when $\dot{M}_\mathrm{ext}<\dot{M}_\mathrm{leak}$ the ring will disperse. On the other hand, as the leaking mass flux is fixed, $\dot{M}_\mathrm{ext}>\dot{M}_\mathrm{leak}$ signifies that there is an excess mass flux, which in the CRM is converted into planetesimals. We see the expression \eq{Mleak} is independent of external mass flux. Therefore, once a ring is present, the mass flux downstream is set and will not be influenced by any fluctuations in $\dot{M}_\mathrm{ext}$ upstream.

The analytical values calculated basing on \eq{Mleak} are consistent with the numerical results in magnitude ($f_\mathrm{edg}\simeq 1$), but the leaking mass flux in runs where clumps don't drift is only half of the cases where clumps with advection. The difference between these two sets tells that clump advection increases $\dot{M}_\mathrm{leak}$ by contributing to the leaking. Small $\alpha_\mathrm{c}$ clump will not affect as we expect in \eq{Mleak}, but in run \texttt{t1f2a4} \hcadd{(see \se{ringmorph})}, we see the leaking mass flux is larger than \texttt{t1f2} \hcadd{(see \se{ringmig})} although they are different only on $\alpha_\mathrm{c}$. This is because clump radial diffusion cannot be ignored with such a high $\alpha_\mathrm{c}$. Similar to clump advection, radial diffusion of clump promotes the leaking mass flux as well.

\begin{figure}
    \centering
    \includegraphics[width=0.99\columnwidth]{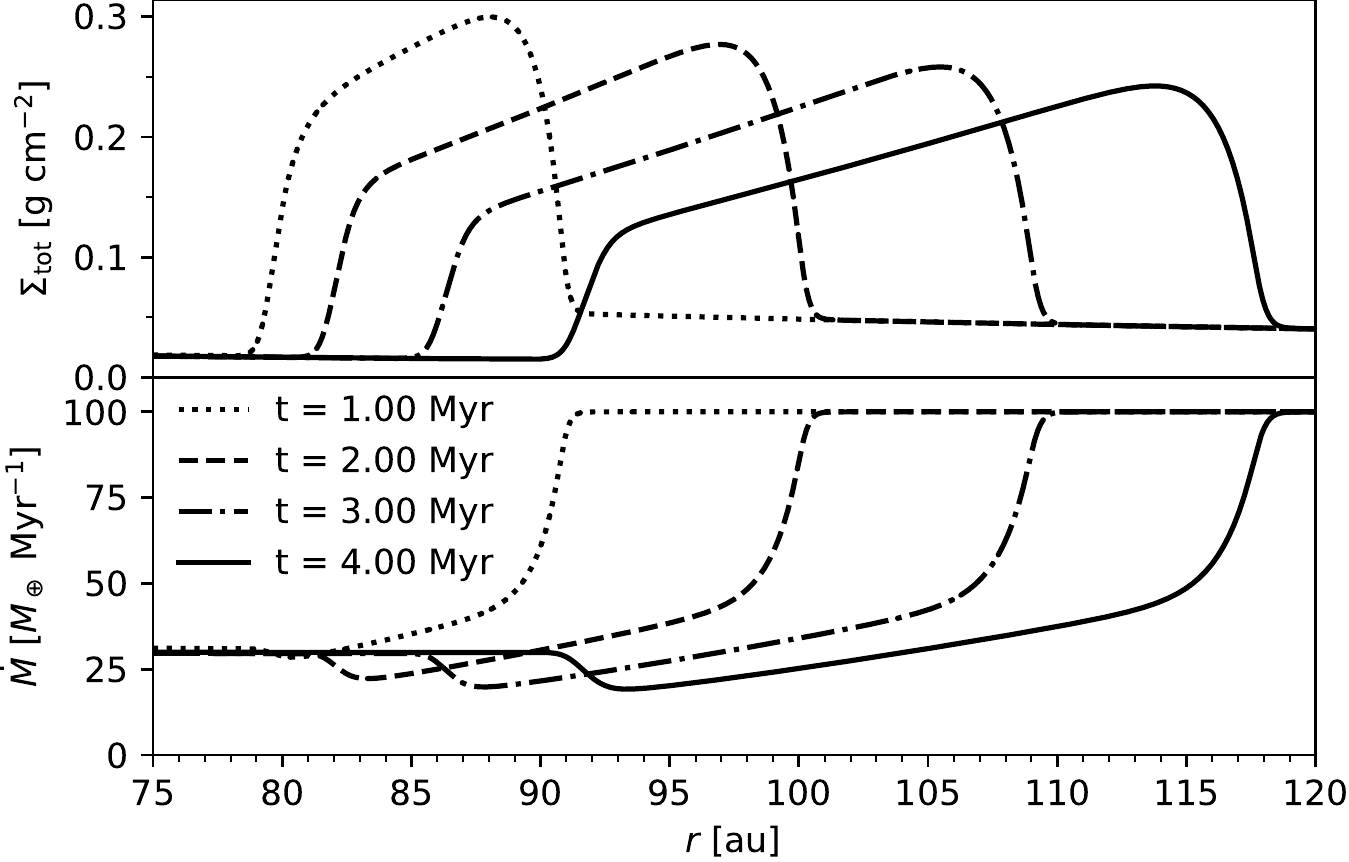}
    \caption{\label{fig:t1f2-evo}Run \texttt{t1f2}. Same as \Fg{t1f1-evo} but no clump advection. Ring profile moves faster than that in default model \texttt{t1f1}. Parameters are listed in \Tb{models}.}
\end{figure}

\subsection{Ring Migration}\label{sec:ringmig}
As can be seen from the mass flux panel of \fg{t1f1-factor} (middle panel), the clump's mass flux is overall positive, indicating that clump pebbles are drifting towards the host star. However, rather than drifting inward, the ring profile actually moves outwards as shown in \fg{t1f1-evo}. This tells us that the motion of the ring is not an overall migration of the dust inside the ring, but a phase velocity of the morphology. In \fg{t1f1-factor} we find that there is a small negative feature in the mass flux at the outer edge of the ring, which corresponds to the outer edge of the ring. At this point, the radial diffusion of clumps is stronger than its inward drift. These clumps move outward and will grow by feeding from the incoming dust flow. We anticipate the ring to move outward faster when the radial diffusion of the clump is stronger (higher $\alpha_\mathrm{c}$) or the radial drift of disc pebbles is lower. The former helps with the outward diffusion of clumps at the ring edge. The latter weakens the inward drifting to render the outward moving more significant. 

\hcadd{In order to investigate how the clump drift prescription influences the ring's movement, we take several runs with different options for the clump's advection (see \se{transportmodel}). In runs \texttt{t1f1}, \texttt{t1f1c9}, \texttt{t1f1c6}, \texttt{t1f1c3} and \texttt{t1f1c0},} we vary $f_\mathrm{ca}$ from $1$ to $0$ to increase the clump drift velocity, and measure the ring migration velocity if there is ring survive. When $f_\mathrm{ca}$ is lower than 0.3, the initial seed ring does not feed from the external mass flux and decays. For $f_\mathrm{ca}=0.6$ the radial drift of the clumps has become $\sim\!5$ times that of the NSH solution. A ring forms, but due to the high advective velocity it moves inward rather than outward.

\begin{figure}
    \centering
    \includegraphics[width=0.99\columnwidth]{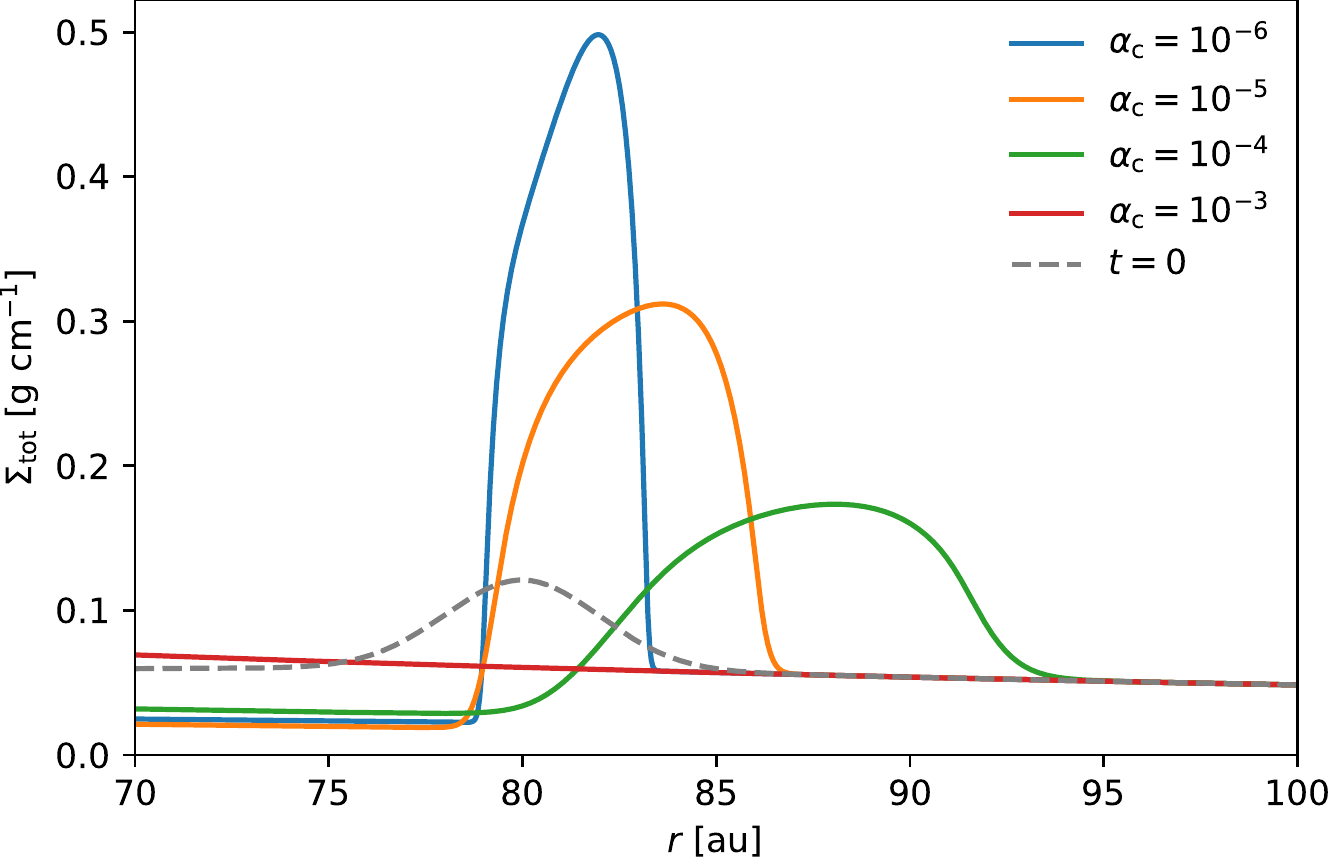}
    \caption{\label{fig:ring-morph} Snapshots of rings at $t=0.5$ Myr, varying the clump diffusivity $\alpha_\mathrm{c}$. The Gray line is the initial pebble distribution for all of the three runs. Higher $\alpha_\mathrm{c}$ broaden the rings and result in stronger outward motion, until at $\alpha_\mathrm{c}=10^{-3}$ the ring no longer survives.}
\end{figure}

Next, we consider the case without any clump advection. Another reason to study this $v_\mathrm{dr,c}=0$ case is that it allows us to study the outward ring motion in isolation. Accordingly, we keep all parameters the same as in run \texttt{t1f1}, but switch off the advection term. The results of this simulation (\texttt{t1f2}) is shown in \fg{t1f2-evo}. Similar to run \texttt{t1f1}, within $1$ Myr stage the location of the ring peak did not change but the height grows. On the other hand, after $1$ Myr a massive ring forms with the front edge of the ring moving much faster than in run \texttt{t1f1}. In this case, the pileup of clumps at the outer edge happens more easily, because the clumps which accumulate at the ring region do not drift inward.

\begin{figure}
    \centering
    \includegraphics[width=0.99\columnwidth]{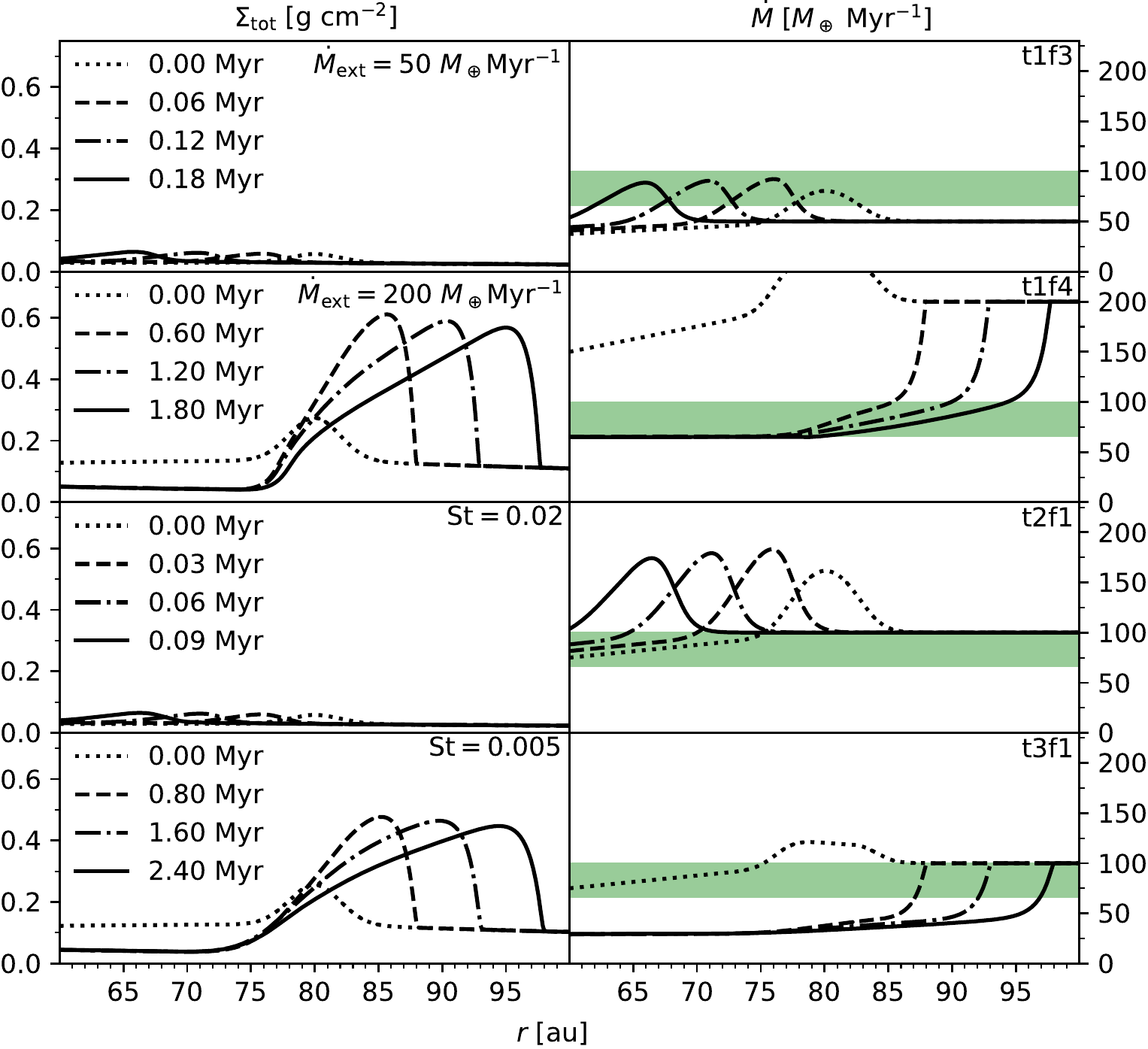}
    \caption{\label{fig:t1f1-StM} Ring survival depends on Stokes number and mass influx rate. From up to down, run \texttt{t1f3}, \texttt{t1f4}, \texttt{t2f1} and \texttt{t3f1}. In the upper two panels, the external mass flux is lower ($50 M_\oplus$ Myr$^{-1}$) and higher ($200 M_\oplus$ Myr$^{-1}$) compared with the the default model \texttt{t1f1}, while in the lower two panels the Stokes number is larger ($0.02$) and smaller ($0.005$). The green bar's upper side indicates the external mass flux, and lower side of green bar indicates the leaking mass flux in the default model. In order to preserve the ring, the external mass flux must be larger than the leaking mass flux. Therefore, larger $\dot{M}_\mathrm{ext}$ and lower St aid the long-term survival of the ring.}
\end{figure}

\begin{figure}
    \centering
    \includegraphics[width=0.99\columnwidth]{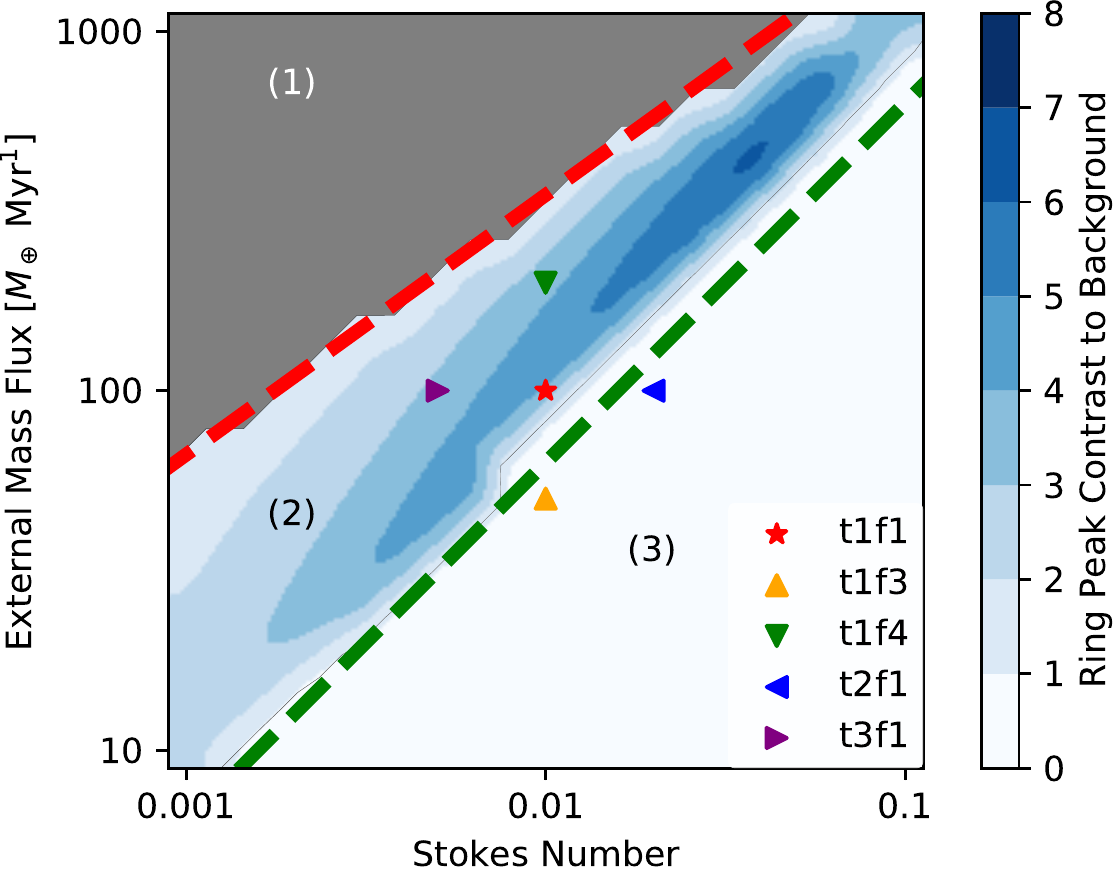}
    \caption{\label{fig:para1}Parameter study about the survival of the ring. \textbf{Gray}: Regions where the initial background dust-to-gas (without a ring) is larger than $0.5$ (indicated by the red line). Excluded from our analysis. \textbf{Blue}: Regions where rings form. The colour indicates the ring surface density-to-background ratio (contrast); \textbf{White}: Regions where the initial perturbation has decayed. The green dashed line corresponds to the leaking mass flux for each St. When $\mathrm{St}=0.04$ and $\dot{M}_\mathrm{ext}=400\,\mathrm{M}_\oplus\,\mathrm{Myr}^{-1}$ maximum contrast is obtained.}
\end{figure}

We estimate the magnitude of the ring migration as follows. Because of the sharp gradient at the outer edge of the ring, clumps diffuse outward radially. After their diffusion, clumps will be transformed to disk pebbles on a vertical diffusion time-scale $t_\mathrm{diff} = H_\mathrm{d}^2/D_z = t_\mathrm{sett} = 1/\mathrm{St}\Omega_K$. At the same time, these clumps at the front edge of ring have spread radially over a region $\Delta x = \sqrt{D_\mathrm{c}t_\mathrm{diff}}$ before they disperse. The upstream-scattered clumps intercept a fraction $P$ of the disc pebble mass flux, where $P = f_\mathrm{acc}(\Delta x/v_\mathrm{dr,d}) /t_\mathrm{sett}$ is the ratio of the time pebbles drift over the region $\Delta x$ and the time it takes them to settle into the clumpy sublayer. Considering the "accretion efficiency", we leave an order of unity factor $f_\mathrm{acc}$. Therefore, the accumulation of clumpy material at the front edge of the ring progresses at a rate:
\begin{equation}\label{eq:clump_acc}
    \left. \dfrac{\mathrm{d} M}{\mathrm{d} t} \right\vert_\mathrm{front}
    = P \dot{M}_\mathrm{ext} = f_\mathrm{acc}\sqrt{\mathrm{St}D_\mathrm{c}\Omega_K}\frac{\dot{M}_\mathrm{ext}}{v_\mathrm{dr,d}}.
\end{equation}
Equal the outwards mass flux of \eq{clump_acc} to $2\pi r \Sigma_\mathrm{c} v_\mathrm{ring}$, we obtain the ring migration velocity
\begin{equation}\label{eq:v_ring}
    v_\mathrm{ring} = f_\mathrm{acc}\sqrt{\mathrm{St}D_\mathrm{c}\Omega_K}\frac{\dot{M}_\mathrm{ext}}{2\pi r v_\mathrm{dr,d} \Sigma_\mathrm{c}}
\end{equation}
The clump surface density at the ring front $\Sigma_\mathrm{c}$ can be estimated realizing that the clumping depth of clump $\tau_\mathrm{c}\approx1$. This indicates the transition from clumpy ring region to background where there is no clumps. In the Epstein regime (\Eq{St})
\begin{equation}
    \tau_\mathrm{c} = \kappa \Sigma_\mathrm{c} \sim \frac{\Sigma_\mathrm{c}}{\rho_\bullet a_\bullet} \sim \frac{\Sigma_\mathrm{c}}{\mathrm{St} \Sigma_\mathrm{g}}
\end{equation}
Therefore, $\Sigma_\mathrm{c}\sim \mathrm{St}\Sigma_\mathrm{g}$ and with further substitutions for $D_\mathrm{c}$ and $v_\mathrm{dr,d}$ we obtain
\begin{equation}
\begin{aligned}
    \label{eq:vring}
    v_\mathrm{ring} 
    \simeq & 20\ \mathrm{au\ Myr}^{-1}f_\mathrm{acc} \left(\frac{\alpha_\mathrm{c}}{10^{-5}}\right)^{0.5} \left(\frac{\mathrm{St}}{10^{-2}}\right)^{-1.5} \left(\frac{h_0}{0.075}\right)^{-1} \\
    & \times \left(\frac{\Sigma_\mathrm{g,0}}{3 \mathrm{g\ cm}^{-2}}\right)^{-1} \left(\frac{\dot{M}_\mathrm{ext}}{100 M_\oplus \mathrm{Myr}^{-1}}\right) \left(\frac{r}{100 \,\mathrm{au}}\right)^{-0.25}.
\end{aligned}
\end{equation}
Using this equation, we successfully retrieve the migration velocity of the ring in \fg{t1f2-evo}, which is $9.5\,\mathrm{au}\ \mathrm{Myr}^{-1}$ at $r = 90$ au, with $f_\mathrm{acc}=0.5$. We therefore take $f_\mathrm{acc}=0.5$ for all of our simulation and calculate the estimated values of the ring velocity basing on \eq{vring}, which are listed in \tb{models} as $v_\mathrm{ring}$. In addition, \Tb{models} lists the measured velocity $v_\mathrm{ring,ms}$, which is obtained from the position of the leading edge of clump surface density.

\begin{figure*}
    \centering
    \includegraphics[width=0.99\columnwidth]{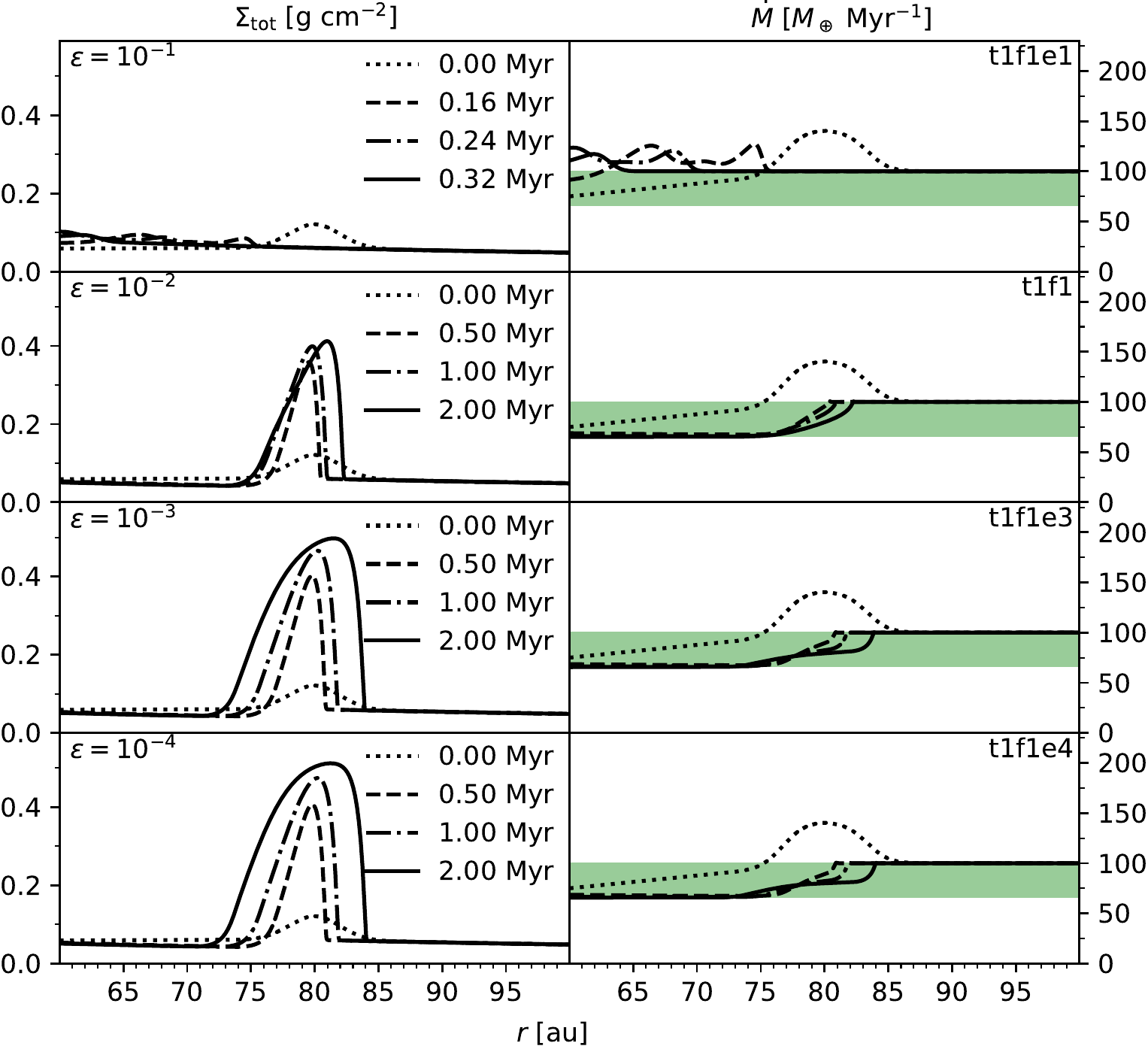}
    \includegraphics[width=0.99\columnwidth]{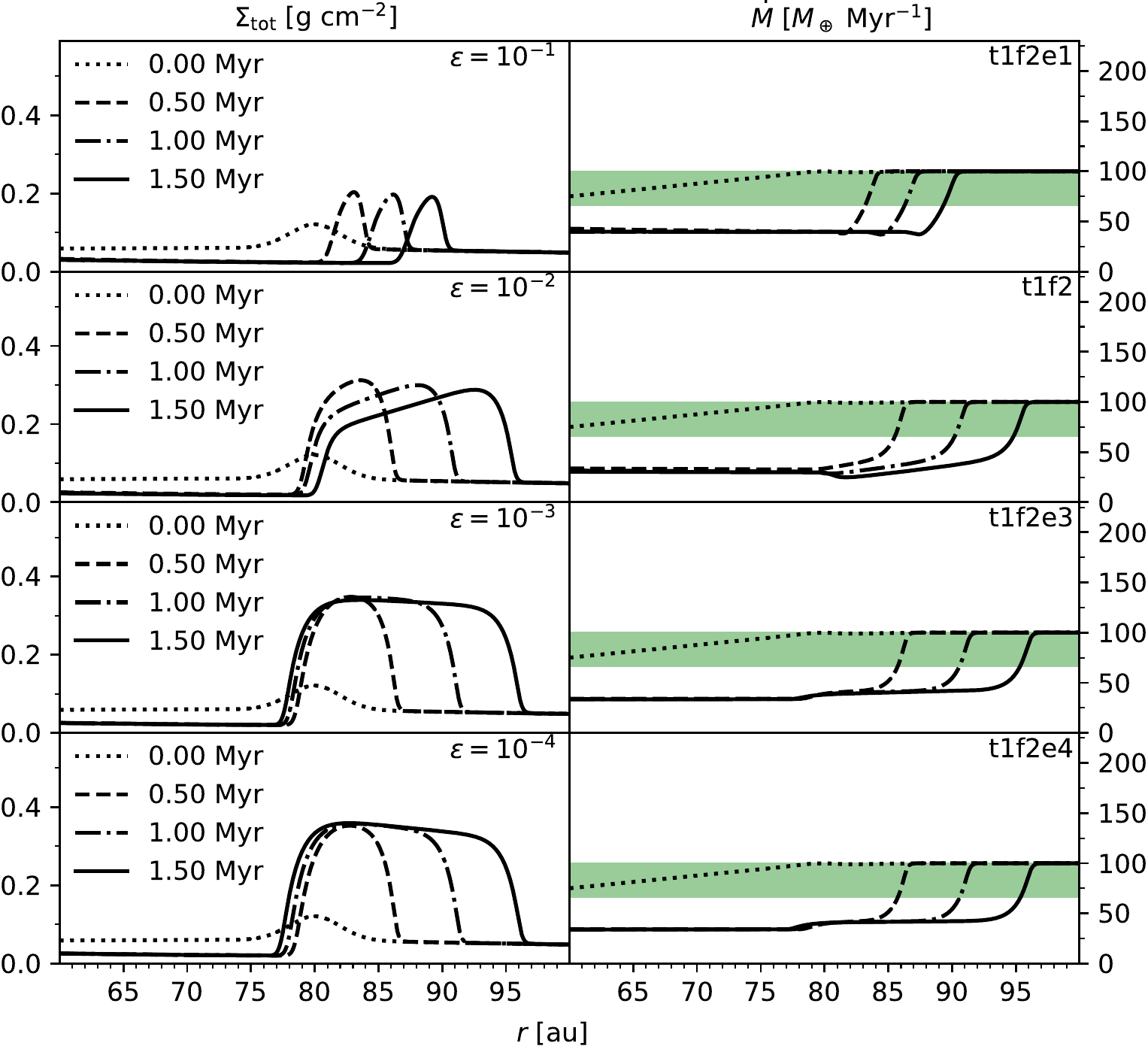}
    \caption{\label{fig:t1f1-eps}Effects of planetesimal formation rate on the outcome of the ring. From top-down, planetesimal formation efficiency $\epsilon=10^{-1}$, $10^{-2}$ to $10^{-3}$ and $10^{-4}$. \textbf{left:} runs with clump advection (group \texttt{t1f1}). Higher $\epsilon$ causes the ring to collapse into planetesimal more rapidly. \textbf{right:} runs without clump advection (group \texttt{t1f2}). Higher $\epsilon$ will significantly weaken the width of the ring.}
\end{figure*}

Comparing the movement of rings in \fg{t1f1-evo} and \fg{t1f2-evo}, one notes that the two ring velocities are different (\texttt{t1f1}: $1.3\,\mathrm{au}\,\mathrm{Myr}^{-1}$, \texttt{t1f2}: $9.5\,\mathrm{au}\,\mathrm{Myr}^{-1}$; measured velocity $v_\mathrm{ring,ms}$, see \tb{models}). The ring in the non-clump-advection run \texttt{t1f2} moves much faster than the ring in the standard model \texttt{t1f1}. The reason is simply that clumps in \texttt{t1f1} have a non-zero radial drift velocity $v_\mathrm{dr,c}$. Thus, the ring itself drifts inward as well. A more accurate prediction is therefore $\sim\!(v_\mathrm{ring} - v_\mathrm{dr,c})$ where $v_\mathrm{dr,c}$ is the clump drift velocity. Since the clump radial drift depends on dust concentration $Z$, which varies with disk radius, a precise analytical expression is hard to obtain, however. But crudely, with $v_\mathrm{dr,c}\simeq\!9\,\mathrm{au}\,\mathrm{Myr}^{-1}$ (individual radial drift velocity when $Z_\mathrm{mid} = 0.5$ in \eq{vdr}), the net velocity of the ring amounts to $\simeq\!1\,\mathrm{au}/\mathrm{Myr}$ for run \texttt{t1f1} which approximately correspond to the measured value. Given the fact that $v_\mathrm{ring}$ and $v_\mathrm{dr,c}$ are of the same magnitude, the net ring velocity and sign become hard to predict. For example, in run \texttt{t1f1h8}, with a higher aspect ratio and faster dust drifting, the clump advection dominates over the ring's outward movement. The ring moves inward in this case \hcadd{(see \texttt{t1f1h6}, \texttt{t1f1h9}, \texttt{t1f2h6}, \texttt{t1f2h8}, \texttt{t1f2h9} for more comparison in \Tb{models}).} For the cases where clumps do not drift, $v_\mathrm{dr,c} = 0$, the predicted ring velocity $v_\mathrm{ring}$ is more consistent with the measured velocity. And for the runs where the clump drift is faster than the NSH solution ($f_\mathrm{ca}<1$), the ring's movement is dominated by the clump inward advective motion.

\Eq{vring} shows that the ring velocity increases with higher clump diffusivity $\alpha_\mathrm{c}$ and higher external mass flux, \hcch{and}{but that the ring migrates faster with decreasing} Stokes number. A higher $\alpha_\mathrm{c}$ helps with the diffusion of clumps; clumps spread faster at the front edge of the ring and speed up its migration as in run \texttt{t1f2a4}. However, further increasing the $\alpha_\mathrm{c}$ is detrimental to the ring survival. In run \texttt{t1f2a3}, $\alpha_\mathrm{c} = 10^{-3}$ and no ring can survive due to the fast spreading of dust. Besides, a higher external mass flux supplies more materials to feed the ring, which leads to more dust mass pileup and then faster migration as well. The Stokes number influences the migration speed significantly. Larger St means that the clump radial drift is faster, which is bad for dust pileup and also tells that the interaction time-scale available for material exchanging is shorter.

\subsection{Ring Morphology}\label{sec:ringmorph}
The CRM produces rings that generally are asymmetric: the outer edge of the ring is usually sharper than the inner edge. Several parameters will change the ring morphology, but the most important one is the clump diffusivity $\alpha_\mathrm{c}$. \Fg{ring-morph} shows snapshots of runs \texttt{t1f2a6}, \texttt{t1f2}, \texttt{t1f2a4} and \texttt{t1f2a3}, where $\alpha_\mathrm{c}$ increases from $10^{-6}$ (10x less than the default) to $10^{-3}$ (100x higher than the default), at $t=0.5$ Myr. As we discussed in \se{ringmig}, larger $\alpha_\mathrm{c}$ speeds up the outward migration of rings. But it also flattens the ring and makes the overall morphology more Gaussian. In run \texttt{t1f2a3}, the clump diffusivity is too high ($\alpha_c=10^{-3}$) that the ring cannot survive. Clump advection and disc pebble diffusivity change the motion of ring but do not affect its shape.

\subsection{Parameter Study of Ring's Survival}\label{sec:ringsuv}
To investigate the survival and migration of the ring, we study how parameters influence its evolution. As discussed in the previous sections, the two most important parameters are the external mass flux and the Stokes number. The external mass flux, which present the amount of material available to feed the ring, decides the disc pebble density according to \Eq{Sigp0}. The Stokes number controls the drifting and settling time-scale, which determines the mass loss from the ring by advection -- the leaking mass -- and the planetesimal formation rate.

\begin{figure*}
    \centering
    \includegraphics[width=0.99\columnwidth]{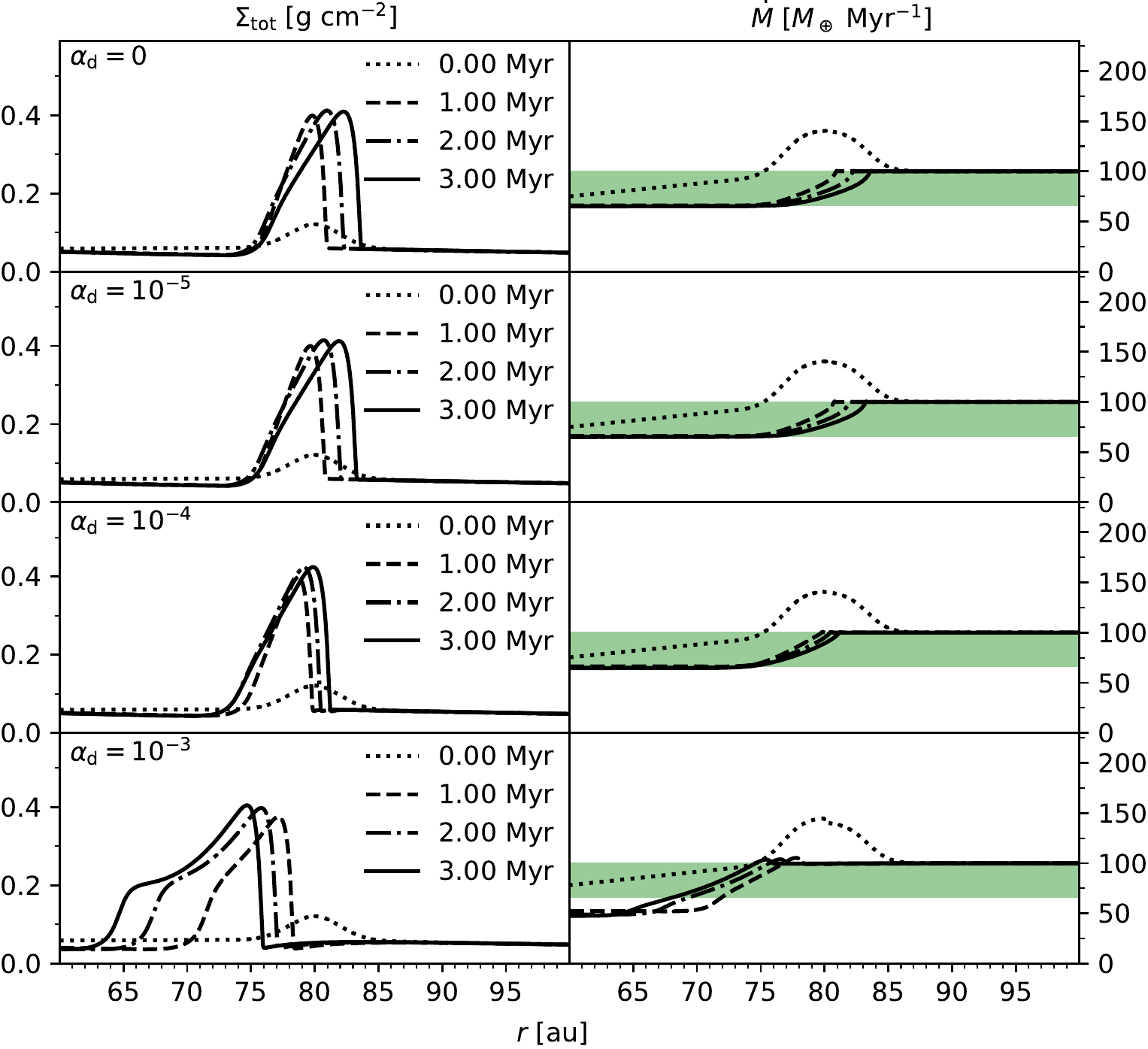}
    \includegraphics[width=0.99\columnwidth]{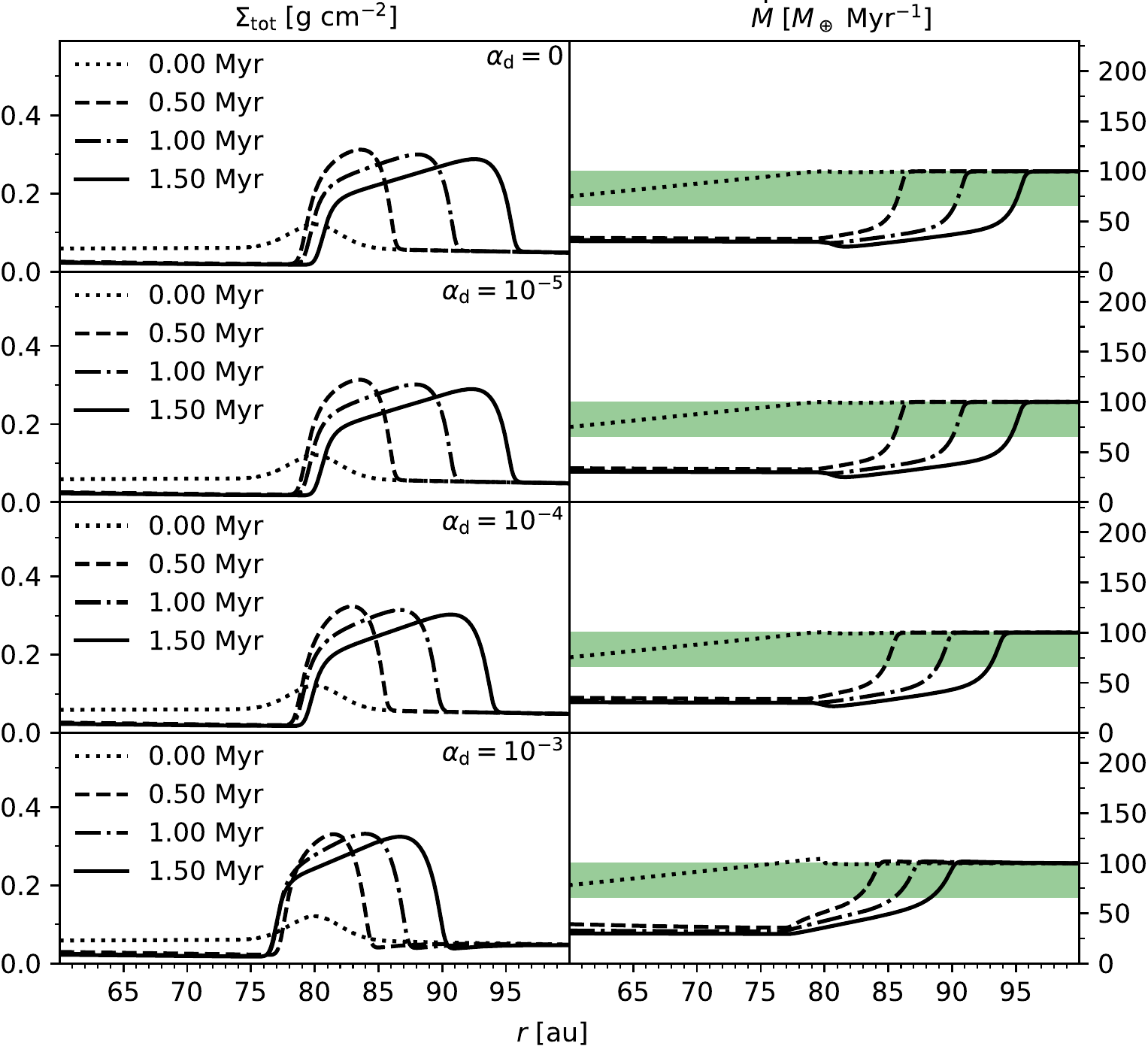}
    \caption{\label{fig:t1f1-n0d}Effects of disc pebble diffusivity on the outcome of the ring. From top-down, disc pebble radial diffusivity coefficient $\alpha_\mathrm{d}$ increase from $0$, $10^{-5}$ to $10^{-4}$ and $10^{-3}$. \textbf{left:} runs with clump advection (group \texttt{t1f1}). Higher $\alpha_\mathrm{d}$ suppress the outward movement of the ring. Ring move inward when $\alpha_\mathrm{d} = 10^{-3}$ \textbf{right:} runs without clump advection (group \texttt{t1f2}). $\alpha_\mathrm{d}$ hardly influence the evolution.}
\end{figure*}

In \fg{t1f1-StM}, we vary the external mass flux. In model \texttt{t1f3}, the constant external mass flux decreases from $100\,\mathrm{M}_\oplus\,\mathrm{Myr}^{-1}$\ to $50\,\mathrm{M}_\oplus\,\mathrm{Myr}^{-1}$ and in model \texttt{t1f4}, it increases to $200\,\mathrm{M}_\oplus\,\mathrm{Myr}^{-1}$(see \fg{t1f1-StM} upper two panels). Since the external mass flux only matters the supplemental mass to the ring, it does not affect the leaking mass. In \texttt{t1f3}, the external mass flux is smaller than the estimated leaking mass flux and the ring disperses because of the lack of dust supply. In contrast, with more material feeding the ring, the ring grows more massive, becomes wider and migrates faster in \texttt{t1f4}. 

In run \texttt{t2f1}, $\mathrm{St} = 0.02$ and both drifting and settling is two times faster than in the default run. As shown in the lower two panels of \fg{t1f1-StM}, the ring disappears very quickly in \texttt{t2f1} with larger St. Larger $\mathrm{St}$ make the initial pebble density lower according to \eq{Sigp0} and also raises the leaking mass flux above the external mass flux. Thus there is no ring. Conversely, for the case $\mathrm{St} = 0.005$ (run \texttt{t3f1}), we get a more massive ring than \texttt{t1f1}.

In \fg{para1} we conduct a parameter study varying the external mass flux ($10 \le \dot{M}_\mathrm{ext} \le 1000\,M_\oplus$) and Stokes number ($0.001 \le \mathrm{St} \le 0.1$). We identify three regions:
\begin{enumerate}
  \item Regions where the background mid-plane dust-to-gas (without a ring) is larger than $0.5$ (gray). We exclude them from our analysis since the pileup conditions are triggered everywhere in the disc, which indicates an inconsistent initial profile.\footnote{Reducing the initial pebble surface density would limit the pileup to the (arbitrary) location of the outer boundary, which is also inconsistent. For runs in the gray region, our simulation setup is inapplicable as clumps (ring formation) is not limited to the ring region. Physically, these regions also correspond to extremely high dust-to-gas surface density ratios ($\gtrsim\!0.3$).}
  \item Regions where rings form. The colour indicates the ring surface density-to-background ratio (contrast) (blue).
  \item Regions where the initial perturbation has decayed within $1$ Myr (white).
\end{enumerate}
From \fg{para1}, it is clear that there is a sharp transition between the ring-surviving blue area (2) and ring decaying white area (3). We draw a green dashed line, which is consistent with the leaking mass flux in \eq{Mleak} with $f_\mathrm{edg}=1.6$. Below this criterion, the leaking mass flux dominates the evolution and rings disperse within $1$ Myr. Above this criterion, the external mass flux is larger than the mass loss due to planetesimal formation and ring leaking at the initial position. The contrast between the ring and background is a key observable. In \fg{para1}, the contrast approaches the maximum around St $=0.04$ and $\dot{M}_\mathrm{ext}=400\,\mathrm{M}_\oplus\,\mathrm{Myr}^{-1}$.

The survival of the rings also depends on the planetesimal formation efficiency $\epsilon$, since it influences the mass loss of the ring directly. We vary the $\epsilon$ from $10^{-1}$ to $10^{-4}$ in run \texttt{t1f1} and \texttt{t1f2} \hcadd{(run \texttt{t1f1e1}, \texttt{t1f1e3}, \texttt{t1f1e4}, \texttt{t1f2e1}, \texttt{t1f2e3}, \texttt{t1f2e4} in \Tb{models})}. The results are shown in the two sides of \fg{t1f1-eps} separately. In the \texttt{t1f1} group, clumps have an advection velocity and the leaking mass flux is higher than that in the \texttt{t1f2} group. In other words, as the external mass flux is the same in these two cases, the allowed upper limit for mass loss by planetesimal formation is lower in the \texttt{t1f1} group than in the \texttt{t1f2} group. For example, the top panels show that run \texttt{t1f1} is unable to keep the ring structure with $\epsilon = 0.1$ whereas \texttt{t1f2} can. In both groups, decreasing $\epsilon$ from $10^{-2}$ to $10^{-4}$ will not change the ring evolution significantly. This is because for low $\epsilon$ mass loss is dominated by pebble drift rather than planetesimal formation. To summarize, inefficient planetesimal formation does not guarantee ring survival, but too efficient planetesimal formation ($\epsilon \gtrsim 0.1$) will certainly destroy the ring.

Finally, we consider the effect of the disc pebble diffusivity on the ring velocity. Until now $\alpha_\mathrm{d}$ has been put zero for simplicity, because the transport of disc pebble is dominated by advection. We set up two groups of simulations with $\alpha_\mathrm{d}\neq0$ as shown in \fg{t1f1-n0d}. From top to bottom, the values of $\alpha_\mathrm{d}$ increase from $0$, $10^{-5}$, $10^{-4}$ to $10^{-3}$. As the disc pebble diffusivity increases, the ring's outwards motion slows down. In the group where clumps can drift, the direction of migration changes sign at $\alpha_\mathrm{d}=10^{-3}$, which is the same as $\delta_\mathrm{d}$ but much higher than the clump diffusivity $\alpha_\mathrm{c}=10^{-5}$. Similarly, in the no-clump advection group (right panels) the rate of ring migration slows down. This reduction of the outward migration can be understood as follows. According to the CRM in \se{ringmig}, the migration of the ring is a result of disc pebbles interacting with outward-diffusing clumps. As shown in \Fg{t1f1-factor}, in the CRM, the disc pebble surface density is lower inside the ring than outside because pebbles are getting incorporated into clumps (and have difficulty to escape). Accounting for a disk diffusivity, a diffusive flux directed towards the minimum of $\Sigma_\mathrm{d}$ -- i.e., the ring location -- arises. This effectively increases the radial motion of disc pebbles at the leading edge, and thus decreases the probability of pebbles to be absorbed into the clumps. Therefore, the clump's inward drift velocity $v_\mathrm{dr,c}$ dominates over the outward $v_\mathrm{ring}$, resulting in a net inward movement of the ring. Rings still move outward in the case without clump advection, albeit at a slower pace.

\subsection{Constant Size Runs: Cyclical Rings}\label{sec:cnst-size}
Another choice for the pebble properties in the simulations is to fix the size of the pebbles, which is consistent with particle growth limited by bouncing barrier \citep{GuettlerEtal2010,ZsomEtal2010}.

\begin{figure}
    \centering
    \includegraphics[width=0.99\columnwidth]{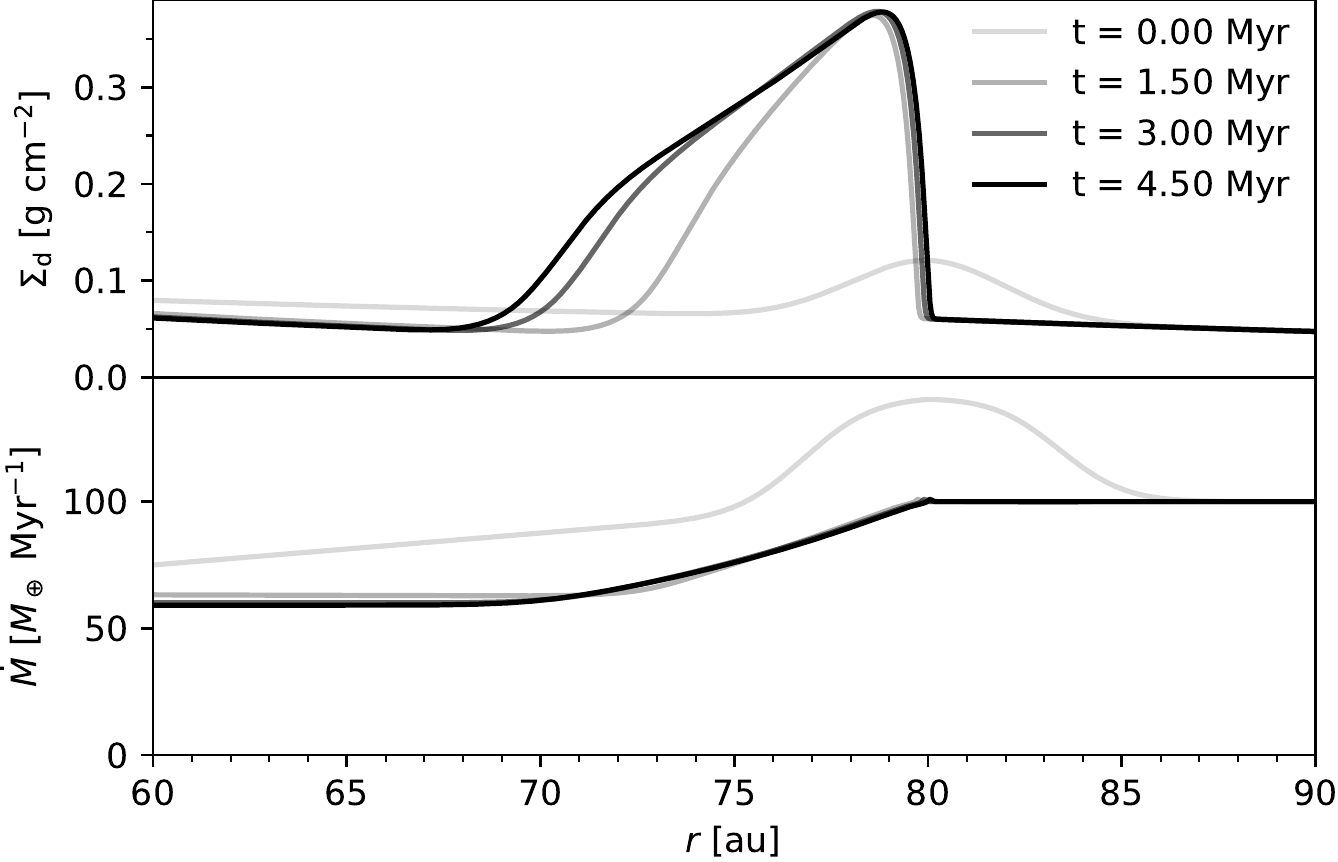}
    \caption{\label{fig:s1f1-zoom}Same as \Fg{t1f1-evo} but with fixed particle size. Parameters are listed in \Tb{models} as \texttt{s1f1}. Clumps that diffuse outward drift back more quickly than \texttt{t1f1} because of the increasing $\mathrm{St}$ at larger radius.}
\end{figure}

In simulation \texttt{s1f1}, we keep all parameters identical to the default run \texttt{t1f1}. The only difference is that this model characterized by a fixed particle radius of $a_\bullet = 103\,\mu$m, which corresponds to a Stokes number of 0.01 at 80 au. The result of this \texttt{s1f1} run is shown in \fg{s1f1-zoom}. Unlike the constant $\mathrm{St}$ case, the front of the ring features no significant outwards movement. We attribute this effect due the increase of the Stokes number with radius: clumps diffuse outward drift back more quickly because of the increasing $\mathrm{St}$ at larger radius. In addition, lower $\mathrm{St}$ at smaller radius means that the leaking mass flux is also smaller. Both these effects help the stability of the ring.

According to \eq{St}, the Stokes number decreases as dust drifts from the outer disc towards the host star, which promotes the traffic jam of drifting pebbles at the inner disc \citep{YoudinShu2002,YoudinChiang2004,DrazkowskaEtal2016}. With the same parameters as in run \texttt{s1f1}, we set up run \texttt{s1f1A0} in which the surface density is a smooth power-law profile at $t=0$. Even without an initial perturbation $(A=0)$, we find that a pileup of pebbles happens in the inner region. After their formation, the ring moves outward as shown in the space-time plot of \Fg{s1f1-coler}. 

As was discussed in the motion of ring part (see \Se{ringmig}), the ring's migration is faster for larger $\mathrm{St}$ and higher $\alpha_\mathrm{c}$. With \hcadd{a higher external mass flux of $\dot{M}_\mathrm{ext} = 400\,M_\oplus\,\mathrm{Myr}^{-1}$, larger fixed particle size} $a_\bullet = 1$ mm and \hcadd{higher clump diffusivity} $\alpha_\mathrm{c} = 10^{-4}$, we get a cyclical process \hcadd{run \texttt{s2f2}} with period $\sim 0.3$ Myr (see \fg{s1f2-coler}). As $\mathrm{St}$ increases with disc radius, the leaking mass flux also increases. The enhanced particle flux leaking from the primary ring towards the secondary ring can form a significant pileup of pebbles at low $\mathrm{St}$ in the inner disc, which supports this cycling. The cycling period get shorter with larger $\mathrm{St}$, higher $\alpha_\mathrm{c}$ and location closer to the star. The physical reason behind this cyclical process can be understood as follows. According to the definition of Stokes number in \eq{St}, $\mathrm{St}$ increases with disc radius along the ring moving outward. Thus the leaking mass flux also increases. An enhanced particle flux is released from the primary ring towards the secondary ring when the primary ring decays, which can assist the secondary ring's external mass flux as well. Consequently, a pileup of material naturally happens in inner disc again, which supports this cycling. 

\begin{figure}
    \centering
    \includegraphics[width=0.99\columnwidth]{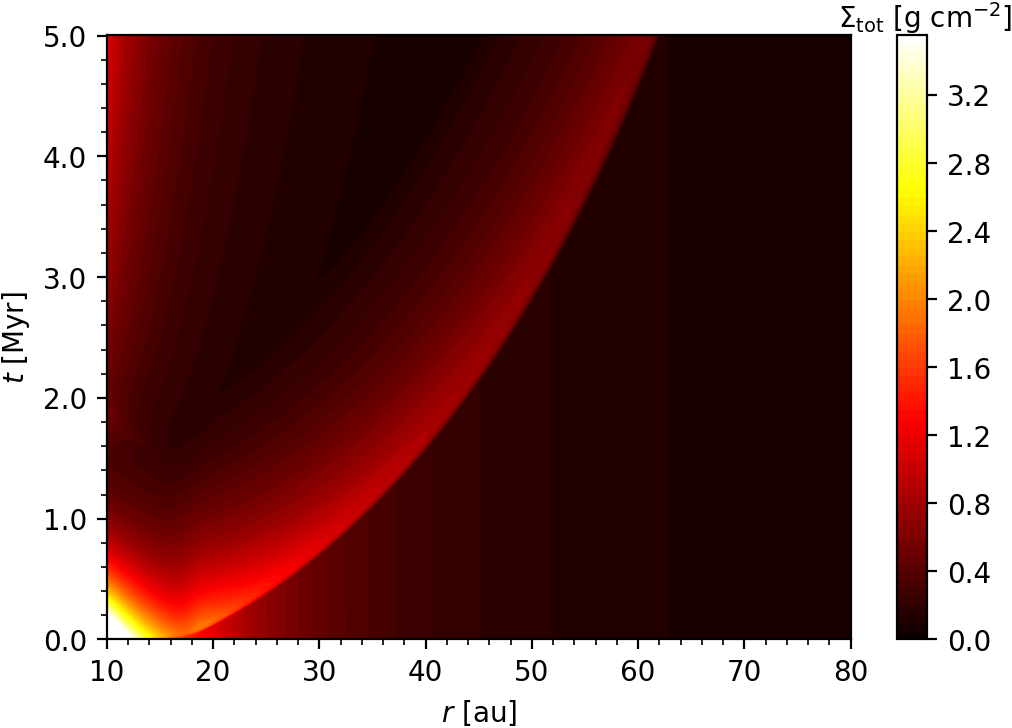}
    \caption{\label{fig:s1f1-coler} Space-time of surface density, illustrating spontaneous ring formation and outward migration (run \texttt{s1f1A0}). Particle size is held fixed in this simulation and there is no initial perturbation. Pileup of ring happens at inner disc region around $10$ au and the ring moves outward to $60$ au.}
\end{figure}

\begin{figure}
    \centering
    \includegraphics[width=0.99\columnwidth]{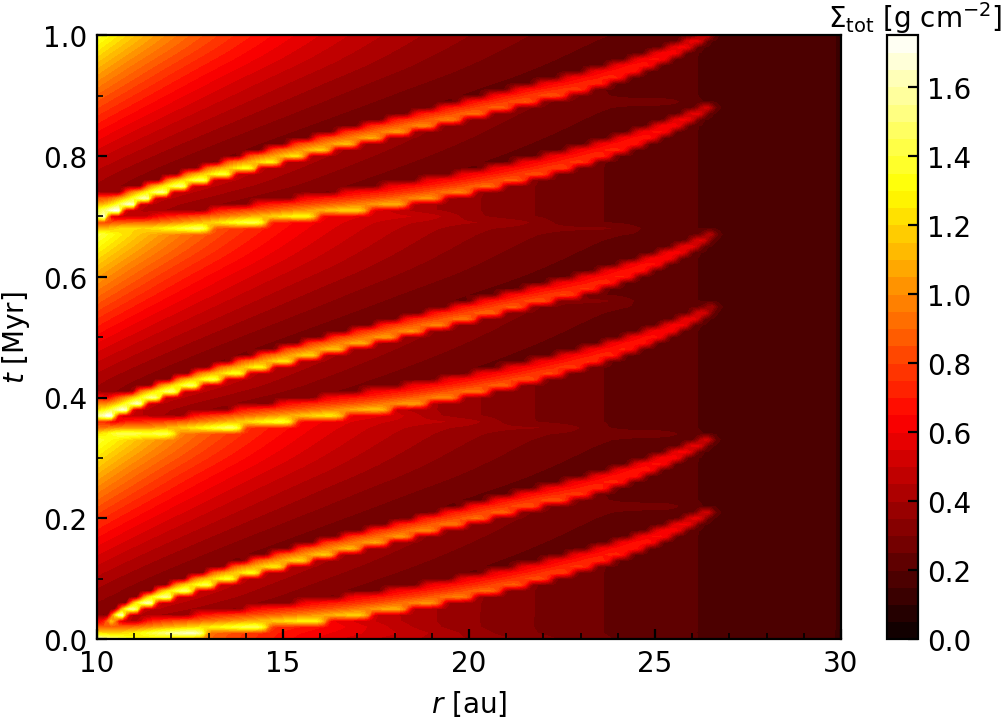}
    \caption{\label{fig:s1f2-coler} Cycling of multiple rings (run \texttt{s2f2}). The evolution of the ring is periodic. Ring is formed by pileup at inner disc and move outward.}
\end{figure}

\begin{figure}
    \centering
    \includegraphics[width=0.99\columnwidth]{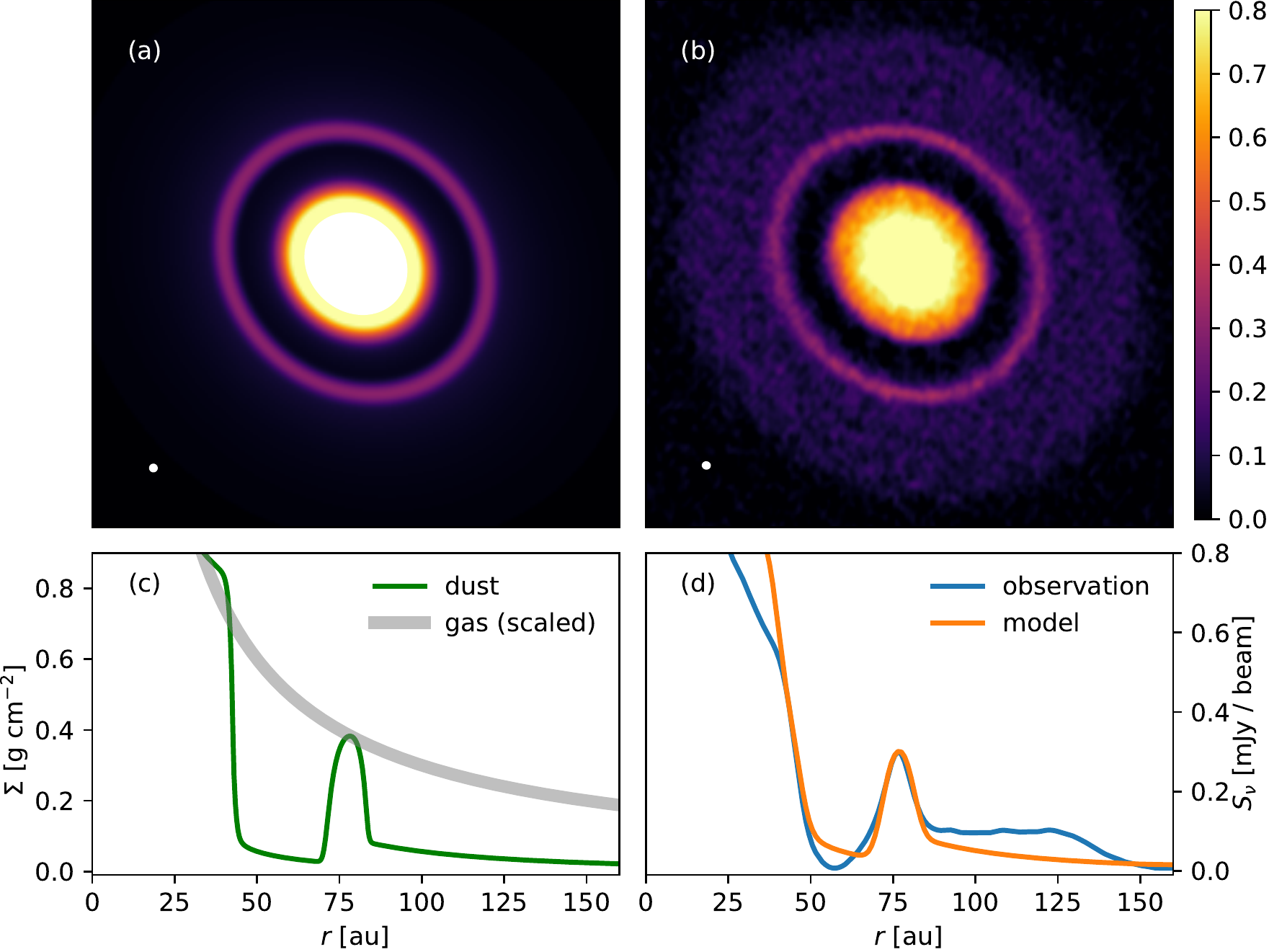}
    \caption{\label{fig:E24} Application of the CRM to Elias 24. \texttt{Panels:} (a) Convolved image resulting from our simulation with constant particle size, used 2D Gaussian beam is plotted at bottom left. (b) Observed continuum emission at $239$ GHz, beam sizes is shown in the lower left. \citep{AndrewsEtal2018}. (c) Total dust surface density (green) and scaled gas density profile (gray). (d) Azimuthally averaged intensity profile from the observation (blue) and intensity profile of the our model (orange).}
\end{figure}

\section{Comparison with ALMA}\label{sec:ALMA}
In order to investigate the spatial dust distribution responsible for the ring emission, we generate a synthetic intensity map corresponding to the CRM output. The procedure is the following:

Starting from the luminosity of the host star $L_\star$, the dust temperature $T_d$ can be estimated by the irradiated flaring disc model
\begin{equation}
    T_\mathrm{d}(r)= \left(\dfrac{\phi L_\star}{8\pi r^2 \sigma_\mathrm{SB}}\right)^{0.25}
\end{equation}
where $\sigma_\mathrm{SB}$ is the Stefan-Boltzmann constant and $\phi = 0.02$ is the flaring angle \citep{ChiangGoldreich1997,D'AlessioEtal2001}. The aspect ratio of the disc is therefore (\eq{cs} and \eq{Td}):
\begin{equation}\label{eq:h_T}
    h = \sqrt{\dfrac{k_B T_d r}{\mu m_p G M_\star}}
\end{equation}
which is a function of host stellar mass $M_\star$, stellar luminosity $L_\star$, and radius $r$.

Then we apply the CRM with constant particle size with parameters given in \se{E24} and \se{AS209} and obtain the surface density profile $\Sigma_\mathrm{d}(r)$ and $\Sigma_\mathrm{c}(r)$, from which we obtain the vertical distribution assumption
\begin{equation}
\begin{aligned}
    \rho_\mathrm{d}(z, r) &= \frac{\Sigma_\mathrm{d}}{\sqrt{2\pi}H_\mathrm{d}}\exp{\left(-\dfrac{1}{2}\dfrac{z^2}{H_\mathrm{d}^2}\right)}
    \\
    \rho_\mathrm{c}(z, r) &= \frac{\Sigma_\mathrm{c}}{\sqrt{2\pi}H_\mathrm{c}}\exp{\left(-\dfrac{1}{2}\dfrac{z^2}{H_\mathrm{c}^2}\right)}
\end{aligned}
\end{equation}
We rotate and incline the disc basing on the position angle and inclination from \citet{AndrewsEtal2018} and calculate the intensity. 

For simplify, we assume that the opacity of all dust in our simulation is a constant $\kappa_\nu \sim 1.0\,\mathrm{cm}^2\,\mathrm{g}^{-1}$, corresponding to a grain radius of $a \sim 100\,\mu\mathrm{m}$ at ALMA band 6 ($\lambda = 0.125 \mathrm{cm}$) according to the DSHARP opacity model \citep{BirnstielEtal2018}. Without accounting for scattering, the intensity profile $I_\nu(r)$ by radiative transfer is
\begin{equation}
\begin{aligned}
    \frac{\mathrm{d}  I_\nu(r)}{\mathrm{d}  \tau} &= - I_\nu + B_\nu(T_\mathrm{d})
    \\
    d \tau &= \kappa (\rho_\mathrm{d} + \rho_\mathrm{c}) \mathrm{d} z
\end{aligned}
\end{equation}
where $B_\nu(T_d)$ is the Planck function. Then, we convolve these intensity maps with 2D Gaussian FWHM $\delta \sim 0\farcs035$ ($\sim 5$ au), which correspond to a beam size and distance from the DSHARP data \citep{AndrewsEtal2018}. The used parameters of host star are listed in \Tb{simobs}.

We apply the CRM with constant particle size towards two protoplanet discs, Elias 24 and AS 209 (\fg{E24} and \fg{AS209}). The two sample discs are selected from ALMA DSHARP survey \citep{AndrewsEtal2018,HuangEtal2018,GuzmanEtal2018}. Both of them have well-resolved annular substructures at $239$ GHz dust continuum and they are one of the sharpest rings who hold substantial peak-to-valley contrast among all DSHARP sources \citep{DullemondEtal2018}. We briefly discuss how well the CRM reproduces these substructures. 

\begin{figure}
    \centering
    \includegraphics[width=0.99\columnwidth]{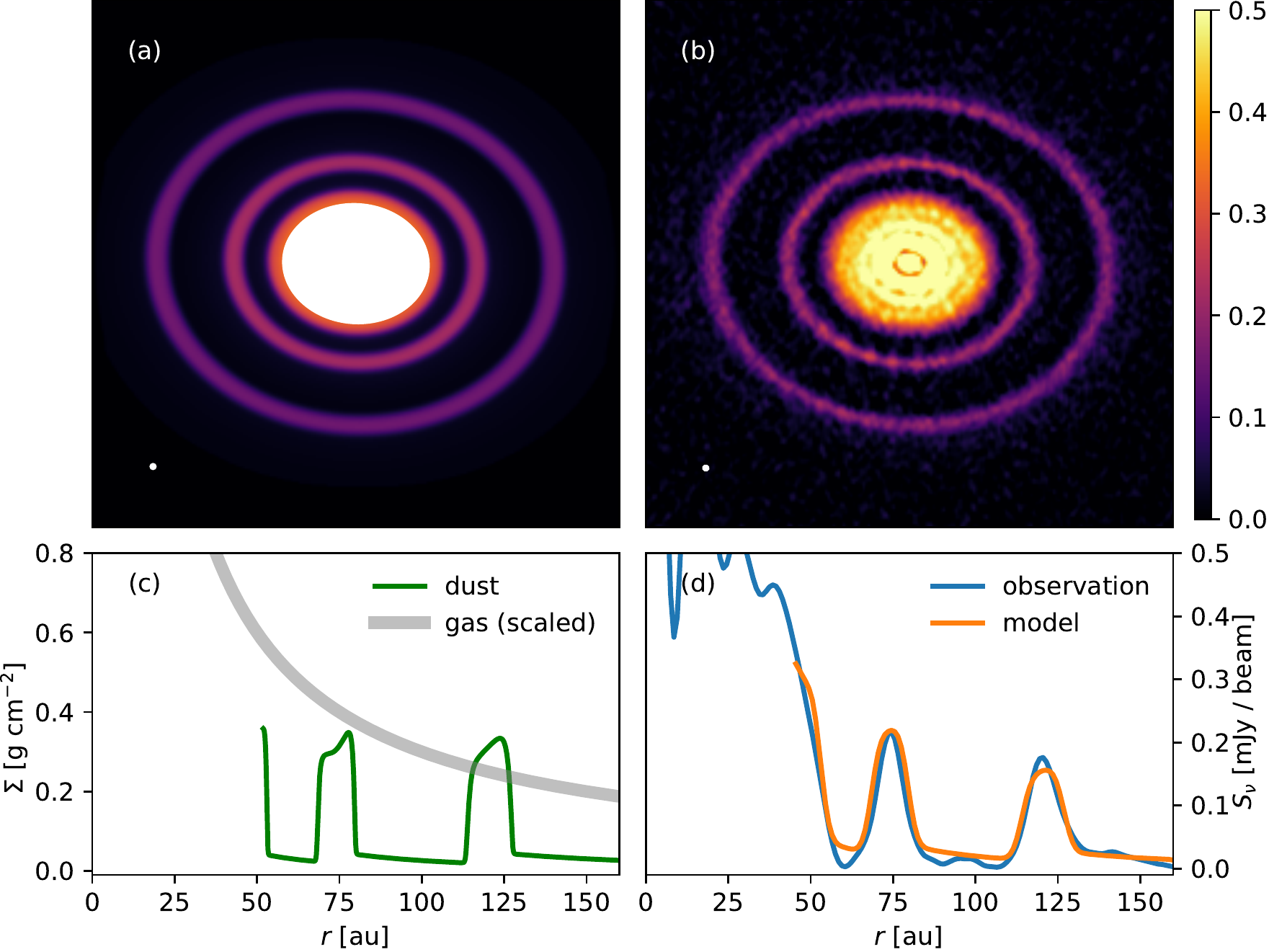}
    \caption{\label{fig:AS209} Application of the CRM to AS 209. \texttt{Panels} same as \fg{E24}}
\end{figure}

\begin{table}
\caption{Stellar parameters and image setups used for simulations \citep{AndrewsEtal2018}.\label{tab:simobs} (1) distance in parsec; (2) (3) stellar luminosity and mass in units of the solar values; (4) beam size used for 2D Gaussian kernel convolution; (5) opacity of dust.}
\centering
\small
\begin{tabular}{l|ccccc}
\hline\hline
name & d & $L_\star$ & $M_\star$ & $\theta_b$ & $\kappa$ \\
& [pc] & [$L_\odot$] & [$M_\odot$] & [mas] & [cm$^2$ g$^{-1}$]\\
& (1) & (2) & (3) & (4) & (5) \\
\hline
Elias 24 & 136 & 0.78 & 6.03 & 35.5 & 1.46\\
AS 209 & 121 & 0.83 & 1.41 & 37 & 1.75 \\
\hline
\end{tabular}
\end{table}

\subsection{Single ring: Elias 24}\label{sec:E24}
First, we show a case featuring a single ring. Elias 24 is a system with a narrow bright ring at $77$ au and a deep gap at $57$ au \citep{HuangEtal2018}. We choose an external mass flux of $\dot{M}_\mathrm{ext} = 200\,M_\oplus\,\mathrm{Myr}^{-1}$, a clump diffusivity $\alpha_\mathrm{c} = 3\times 10^{-5}$ and a particle size of $93.4\,\mathrm{\mu m}$. According to the temperature profile, the aspect ratio $h_0 = 0.1$ at $100$ au. The gas surface density is taken to be $\Sigma_\mathrm{g} =3 \mathrm{g}\ \mathrm{cm}^{-2}\,(r/100 \,\mathrm{au})^{-1}$, hence $\mathrm{St}=8.66\times10^{-3}$ at the ring location ($r=77$ au). We initialize a small perturbation at $r=73$ au with relative amplitude $A=0.5$ (see \eq{sigmatot}).  \hcadd{By $0.45$ Myr the seed ring has transformed into the outward moving state. The ring peak has arrived at the observed location of the ring at submillimeter wavelengths, where} we take a snapshot. In order to scale the intensity, the opacity is chosen to be $\kappa = 1.46\,\mathrm{cm}^2 \mathrm{g}^{-1}$. 

\Fg{E24}{a} shows the convolved continuum map and \Fg{E24}{b} shows the observational dust continuum at ALMA band 6 (239 GHz) and the radial profile, respectively. Panel (c) shows the unconvolved surface density profile corresponding to the simulations and the gas surface density scaled by a factor $0.1$. In panel (d), the orange line shows the azimuthally averaged intensity from the convolved continuum map (a) in orange while the blue line shows the radial profile inferred from the observations \citep{HuangEtal2018}. The continuum emission in panel (b) shows a perfect annular structure and high contrast between the ring to the background. After convolution, the radial profile turns to be smoother and more Gaussian, which make the edges of the ring not as sharp as in the surface density profile. This simulation reproduces not only the positions and intensities but also the shape of the ring and gap well.

\subsection{Two rings: AS 209}\label{sec:AS209}
Most DSHARP disks exhibit multiple annular substructures \citep{HuangEtal2018}. AS 209 is such a system hosting a group of ring \citep{GuzmanEtal2018}. For simplicity, we focus the CRM towards the two most prominent, well-separated outer rings centered at $74$ au and $120$ au. We set up the simulation with parameters $\dot{M}_\mathrm{ext} = 300\,M_\oplus\,\mathrm{Myr}^{-1}$, clump diffusivity $\alpha_\mathrm{c} = 10^{-5}$ and particle size $180\,\mu m$. The aspect ratio $h = 0.08$ at $100$ au correspond to the stellar luminosity of $0.83 L_\odot$ (\eq{h_T}). The gas surface density in power-law profile is $\Sigma_\mathrm{g} =3\ \mathrm{g}\ \mathrm{cm}^{-2}\,(r/100 \,\mathrm{au})^{-1}$, hence $\mathrm{St}=3.75\times10^{-2}$ at the primary ring location $r=120$ au and $\mathrm{St}=2.3\times10^{-2}$ at the secondary ring location $r=74$ au. We initiate two small perturbation with relative amplitude $A=0.2$ at $r=69$ au and with relative amplitude $A=1$ at $r=113$ au. After $0.8$ Myr, we take a snapshot \hcadd{when both seed rings have grown into a steady morphology}. A dust opacity $\kappa = 1.75\,\mathrm{cm}^2 \mathrm{g}^{-1}$ is used to scale the intensity.

The observed dust continuum and radial profile of AS 209 are shown in \fg{AS209} panels (b) and the blue line in (d). Panels (a) and the orange line in (d) are the profiles from the simulation above. The positions, intensities and profile of the rings and gaps are reproduced. The intensity contrast between the two ring is reproduced as well.

The moving ring in the our models show a non-symmetric profile in radial direction, that is, the inner back edge of the ring is always wider and smoother than its outer front edge. New pileup happens at the front rapidly and its motion is opposite to the mass flux direction. Due to the current resolution of observation and the influence of data convolution, the ring is not well enough resolved to estimate the detailed morphology of the ring. But we expect such a shape should be observed in future observation. 

\hcadd{Finally, we briefly discuss how the image changes with the further evolution of the ring. In general, since the simulations have reached a quasi-steady state, the key features in the synthetic image will only slowly change with time. The width of the ring will become a bit wider as the outer edge moves faster than the inner edge of the ring (see \fg{t1f2-evo}). After 1 Myr, for example, the rings in AS 209 will migrate out by a further 10 au, but the intensity and width change by only $\sim\!10\%$. However, it is likely that other model parameters, such as the external mass flux, will change as well on evolutionary timescales, in turn affecting the ring observational characteristics. 
We discuss how the non-constant external mass flux will influence the results in \Se{assessment}.}

\section{Discussion}\label{sec:discussion}
\subsection{Assessment}\label{sec:assessment}
Our scenario to maintain rings on evolutionary time-scales ($\sim$Myr) without a pressure bump relies on the following assumptions: a high mass flux of pebbles ($\sim$100 $M_\oplus\ \mathrm{Myr}^{-1}$), a relatively low but non-zero clump diffusivity ($\alpha_\mathrm{c}\sim10^{-5}$) and a low planetesimal formation efficiency ($\epsilon \lesssim 0.01$). We discuss these parameters and compare them with previous work.

First, the CRM requires a relatively high mass budget for pebbles. The value of the pebble flux in the CRM ($\dot{M}_\mathrm{ext} \gtrsim 100\, M_\oplus\ \mathrm{Myr}^{-1}$) suggests that discs have reservoirs of at least $\sim$100$\,M_\oplus$ in pebbles. These values may seem high compared to typical Class II discs \citep{AndrewsEtal2013,AnsdellEtal2016,BarenfeldEtal2016,AnsdellEtal2017}, but may be more common for the DSHARP sample. 
In addition, the disc dust mass inferred from radio continuum emission could be underestimated because of dust scattering \citep{ZhuEtal2019}. Also, it has been suggested that the disk gas mass -- and by implication the dust mass that "hides" in the outer disk -- is higher than what is inferred from CO observations \citep{PowellEtal2017,PowellEtal2019}. Planet formation models require mass reservoirs of  $\sim$100 $M_\oplus$, as the efficiency of pebble accretion, for example, is low \citep{LiuOrmel2018,OrmelLiu2018}. For example, for the formation of the cores of the solar system's giant planets \citep{LambrechtsJohansen2014} and super-Earth \citep{LambrechtsEtal2019}, typical pebble budgets amount to $130$ and $190$ Earth mass respectively. Thus, it is not unreasonable to anticipate such high mass budgets.

Another simplifying assumption is that the external mass flux is constant. In reality, the mass flux may both fluctuate and and decay with time. The ring will disappear when the mass flux drops below the leaking mass flux $\dot{M}_\mathrm{leak}$ (see \se{Mleak}), as there is not enough material to support the ring any longer. Thus, the lifetime of the ring is determined by the duration over which the external mass flux exceeds $\dot{M}_\mathrm{leak}$. However, the ring is insensitive to small fluctuations in $\dot{M}_\mathrm{ext}$. Even if there was zero external mass flux, rings still takes a time $\sim$$M_\mathrm{ring}/\dot{M}_\mathrm{leak} \approx0.1$--$1$ Myr to disappear. In addition, the sudden release of pebbles may locally enhance the mass flux downstream to create or supply inner rings, see discussion in \se{pb}.

A novel feature of the CRM is the hypothesis that the rings are clumpy in nature, and that these clumps are characterized by a non-zero diffusivity. In (magneto)hydrodynamical simulations of streaming instability a clumpy distribution of solids naturally happen whenever $\rho_\mathrm{d}/\rho_\mathrm{g} \geq 1$ \citep[e.g.,][]{JohansenEtal2006,JohansenYoudin2007,BaiStone2010}. Recently, \citet{YangEtal2017} and \citet{LiEtal2018} have measured the diffusivity of the particles undergoing the streaming stability.  These characteristics -- a clumpy, diffusive medium -- motivate the non-zero $\alpha_\mathrm{c}$ we used. Following the CRM about the clump scale height in \se{clumpmodel}, we express this diffusivity in term of an $\alpha$-viscosity ($D_\mathrm{c} = \alpha_\mathrm{c} H_\mathrm{g}^2 \Omega_K$). Equating the diffusion time-scale of clumps $t_\mathrm{diff} = H_\mathrm{c}^2/D_\mathrm{c}$ with the settling time-scale \eq{tsettle}, we obtain
\begin{equation}
    \delta_\mathrm{c} = \mathrm{St}\frac{H_\mathrm{c}^2}{H_\mathrm{g}^2} \simeq 6 \times 10^{-6} \left(\frac{\mathrm{St}}{10^{-2}}\right) \left(\frac{h_0}{0.075}\right)^2
\end{equation}
\citep[cf.][]{YoudinChiang2004,TakeuchiEtal2012} 
where we, for simplicity, have assumed that the diffusion is isotropic $\alpha_\mathrm{c} = \delta_\mathrm{c}$. For the CRM to produce rings, however, the value of $\alpha_\mathrm{c}$ cannot be too high: when it were to exceed the disc pebble vertical diffusivity $\delta_\mathrm{d}=10^{-3}$, the ring would simply spread out, rendering the mid-plane dust-to-gas ratio to fall below unity. On the other hand, if the clump radial diffusivity is lower than $\alpha_\mathrm{c}=10^{-5}$, the outward ring migration proceeds slower, but it does not affect the survival of the ring. 

Planetesimal formation is another mechanism that influences the survival of the ring \citep{StammlerEtal2019}. It is expected that streaming instability and planetesimal formation happen when the local volume dust-to-gas ratio $>$1 \citep{UmurhanEtal2020} after which we expect the clumps to exist for a settling time-scale \citep{ShariffCuzzi2015}. Hence $\epsilon=1$ implies optimal conversion of (clumpy) pebbles to planetesimals, in which case there would be no ring. Generally, fast planetesimal formation ($\epsilon\sim0.1$) is problematic for \emph{every} planetesimal-forming ring model. For example, in \citet{DrazkowskaEtal2016} $\epsilon \simeq 0.001$ (more precisely, $0.01/2\pi$) is taken as a standard value for $\mathrm{St} = 0.01$ particles, smaller than our standard value of $\epsilon = 0.01$, which corresponds to a time-scale $\gtrsim$1 Myr for planetesimal formation from clumps. Nevertheless, high $\epsilon$ seems in line with hydrodynamical simulations, which typically feature rapid planetesimal formation \citep[e.g.,][]{AbodEtal2019}. However, the ensuing collapse of a bound clump into much more compact planetesimals is under active study with some studies pointing out that turbulent diffusion may suppress the collapse of low-mass clumps \citep{KlahrSchreiber2020}.

Of course, the simplest way to make massive rings for the ALMA rings  is to avoid the conditions for planetesimal formation (streaming instability), i.e., by having a low mid-plane dust-to-gas ratio. But this requires a more turbulent environment with rings likely to become vertically thick (see discussion in \se{pb}).

\subsection{Comparison to Pressure Bump scenario}\label{sec:pb}
In the pressure bump scenario, the width of the ring is determined by the radial diffusivity and the value of the pressure gradient reversal \citep{DullemondEtal2018}. Recently, \citet{RosottiEtal2020} have combined kinematic data (the rotation profile) with the measured ring widths, deriving an aerodynamic constraint of $\alpha/\mathrm{St} \sim 0.1$ for the ring particles, which suggests that the dust radial diffusivity is at least moderate. On the other hand, geometric reasoning suggests a low value of vertical $\delta < 10^{-3}$ as deduced from the pebble scale height $\sim$1 au at $r = 100$ au \citep[e.g.,][]{PinteEtal2016,VillenaveEtal2020}. Several suggestions have been proposed to fix this disagreement. One solution mentioned by \citet{RosottiEtal2020} is that the particles are very small ($\mathrm{St}\lesssim10^{-3}$). Another is to invoke that the diffusivity in the radial and vertical direction is different \citep{ZhuEtal2015,XuEtal2017,YangEtal2018,BaehrZhu2021}. the CRM does not suffer from this tension, as clumps are located within a thin vertical scale height "by definition" \citep{Sekiya1998} and the radial width of rings is independent of $\alpha$.

While the CRM maintains a long-lived ring, it is unclear whether the mechanisms responsible for the pressure bump ring  can be sustained on $\sim$Myr time-scales. \citet{TakiEtal2016} point out that the outward flow of gas created by the back-reaction from dust to gas will flatten the gas pressure bump. Therefore, a mechanism supporting the bump must be present. One such mechanism is the torques exerted on the disc by planets \citep{LinPapaloizou1986}. Previous work by \citet{ZhangEtal2018} suggests that giant planets in the disc can explain annular substructures in some of DSHARP discs \citep{AndrewsEtal2018}. However, in that case one expects a kinematic imprint -- velocity kinks in the rotation curves \citep{TeagueEtal2018,PinteEtal2020}. Recent efforts to obtain kinematic imprints of embedded protoplanet only found convincing detections in 8 out of 18 DSHARP sources \citep{PinteEtal2020}. In particular, there is no obvious CO emission kink in AS 209, despite the high data quality for this source \citep{PinteEtal2020}. Therefore, not every ring is necessarily associated with a planet, justifying exploring alternative mechanisms.

A common feature in the DSHARP observations is the presence of multiple rings (e.g. AS 209, see \se{AS209}). Different from the pressure bump scenario, which traps essentially all of the incoming pebbles, the inward pebble flux is only partially halted by the ring in the CRM. The leaking of inward-drifting pebbles can explain such a multi-ring feature naturally, which is a particularly favored outcome in the constant size model, because the mass flux required to support the ring increases with disc radius (see \se{Mleak}). In other words, although the upstream flux is partly reduced by the outer ring, the remaining flux is still large enough to maintain a second ring or multiple rings downstream.

We discuss several ways to distinguish our scenario from the pressure bump scenario observationally. First, because the CRM does not rely on a pressure maximum, we do not expect ALMA rings to be accompanied with concomitant substructures in gas tracers (such as CO isotopologues). Some early studies to link dust substructure with gas observations are now being made, which is a promising avenue to differentiate between these two scenarios \citep[e.g,][]{KimEtal2020,RosottiEtal2021}. However, interpretation of gas substructure is hard because of the relatively low spatial resolution and sensitivity of gas emission lines and may be further affected by radiation transfer effects. 
Second, the size distribution downstream of the ALMA ring is expected to be different from the distribution inside the ring. In the pressure-bump induced ring scenario, large grains are more efficiently trapped by the pressure bump, whereas the small grains filter through \citep{PinillaEtal2015,PowellEtal2019}. In contrast, in the CRM any particle can leak from the ring regardless of size. Therefore, measuring the change in particle size between the ring location and the region downstream of it is another way to differentiate between the two scenarios. Third, the morphology of the ring in our simulation displays a typical asymmetric profile -- i.e., the leading (outer) edge is steeper than the trailing (inner) edge (see \se{ringmorph}). Higher spatial resolution observation are needed to resolve the ring morphology. For example, the steep "cliffs" edge in AS 209 (\fg{AS209}c) would show up much stronger when the spatial resolution would improve to $\sim\!0\farcs01$.

Finally, it is possible for the two ring mechanism to work in tandem. A pressure bump is certainly a compelling way to concentrate solids and to quickly build rings, particularly in the outer disk. However, pressure bumps may not last for evolutionary time-scales -- planets, for example migrate. The CRM then addresses how rings survive in the long run.

\subsection{Implications for planet formation}\label{sec:outward}
The outward migration of rings seen in most model setups carries important implications for planet formation and debris discs. Strictly speaking, the outward movement of the ring we discuss in \se{ringmig} is only the migration of the leading edge. This outward motion means that the leading edge may not be greatly influenced by (planet formation) processes happening downstream. Pebbles released by the "propagating" ring (as discussed in \se{Mleak}) can aid planet formation in the inner disk. For example, rings could start in the inner disk regions ($\sim$10 au) but then move outwards ($\sim$50 au) in run \texttt{s1f1A0} (\fg{s1f1-coler}). Then in the early stage, the massive ring can help with the formation of planet embryos, while in the later phases it still leaks pebbles to feed the possible inner planet. Therfore, in our scenario outer rings are beneficial towards forming interior planets.

Our outward propagating pebble ring leaves behind a planetesimal belt up to $\sim$100 Earth mass at a surface density $\sim\!1\,\mathrm{g\,cm}^{-2}$. Although these are massive belts, the collision time-scale of planetesimals at $r\approx100$ au, $t_\mathrm{coll}\sim \rho_\bullet R_\mathrm{plts}/\Sigma \Omega$, is still long especially when planetesimals are "born big"  \citep[$R_\mathrm{plts}\sim\mathrm{100\,km}$;][]{MorbidelliNesvorny2020}. 
In other words, we expect migrating rings in the outer disk ($\gtrsim$50 au) to leave behind massive planetesimal belts. Upon dispersal of the disk these planetesimal belts may be stirred to appear as debris disks. Intriguingly, it has recently been suggested that certain debris disks could indeed be very massive. By fitting the observed luminosity to state-of-the-art collision evolution models, one deduces that the primordial planetesimal belts could be as massive as 1000 $M_\oplus$ \citep{KrivovEtal2018,KrivovWyatt2021}! Although these numbers are associated with stars more massive stars than the DSHARP sample, rendering a 1-1 comparison somewhat problematic, the question is raised which mechanism could have created such massive belts. In our view, ALMA rings could well be cradle for such massive planetesimal belts. 

\hcadd{Finally, we briefly discuss the potential of planet formation by pebble accretion in the ring region. In the standard pressure supported disk without pebble clumping, the headwind velocity always contributes to the relative velocity between pebble flow and the planetesimal. Pebble accretion would be ineffective at distances $\sim\!50$--$100$ au, unless the bodies have grown to masses similar to Pluto \citep{VisserOrmel2016}. However, in the CRM the pebble drift is significantly decelerated because of the high dust-to-gas ratios. Therefore, the pebble accretion threshold will be much lower inside the clumpy ring. Thus, in the interval where clump pebbles and planetesimals co-exist, (small) planetesimals could efficiently feed from pebbles. On the other hand, mutual scattering of planetesimals would oppose growth \citep{LiuEtal2019,SchoonenbergEtal2019}. In a future work, we intend to investigate the potential for planet formation in these ALMA rings.} 

\section{Conclusions}\label{sec:conclusions}
In this work, we have developed a model to understand the axisymmetric substructure seen by ALMA in protoplanetary disks -- rings -- without relying on a pressure maximum. This clumpy ring model (CRM) assumes that rings are the manifestation of a dense and clumpy medium, which is actively forming planetesimals. We have conducted a series of  1D transport simulations where pebbles are being characterized either by a fixed Stokes number or by a fixed particle size. In these, we have investigated how key parameters as the pebble size, the external mass flux, and planetesimal formation efficiency affect the survival of the ring. Our main findings are:

\begin{enumerate}
    \item The CRM relies on an initial non-linear perturbation or order unity above the background level. Once it is triggered and the conditions for ring survival are met, a (quasi)-steady state emerges where the ring can exist on evolutionary time-scale.
    \item Without containment (from a pressure maximum), rings leak material at a rate given by \eq{Mleak}. Ring survival demands the existence of an external flux of disc pebbles, exceeding this number. In addition, the time-scale on which planetesimals form must be much longer than the settling time-scale. Once these conditions are met, a ring can be maintained for $\sim$Myr. A higher external mass flux, smaller particle size, and lower aspect ratio facilitate the long-time survival of rings.
    \item A planetesimal formation efficiency lower than the default of $\epsilon=10^{-2}$ does not affect the survival of the ring. On the other hand, efficiencies $\epsilon \gtrsim 0.1$ either destroy or else significantly diminish the extent of the ring.
    \item Due to the diffusive nature of clumpy material, rings move outwards at rates up to $\sim\!10\,\mathrm{au}\,\mathrm{Myr}^{-1}$, while material leaks away from the ring downstream. Outward ring movement is favoured by a higher clump diffusivity (but no more than $a_\mathrm{c} = 10^{-3}$), small clump advection velocity, a larger external mass flux and a lower disc pebble Stokes number.
    \item Compared to the pressure bump ring formation model, our rings stand out for their sharp edges -- a distinction between the CRM and the pressure bump scenario, which can potentially be observed at increased spatial resolution.
    \item Mock images of DSHARP sample Elias 24 and AS 209 are made, which reproduces not only the positions and intensities but also the shape of the ring and gap well.
    \item The high leaking pebble mass flux implies that in the CRM ALMA rings can aid planet formation interior to the observed rings. 
    \item When the particle size is constant throughout the disk, a pileup in the inner disk naturally occurs, which spontaneously triggers the CRM. A cyclical process emerges where rings appear in the inner disk ($\sim$10 au), move outwards over several tens of au before dissolving entirely to re-appear in the inner disk. A constant pebble size also naturally gives rise to multiple rings.
    \item Up to $\sim$$100\,M_\oplus$ in planetesimals are expected to form over a period of $\sim$Myr. In the outer disk, this could be a possible explanation for the high mass budget of planetesimals inferred in debris discs.
\end{enumerate}

To further investigate the viability of the CRM to reproduce ALMA morphology, we plan to develop a more detailed fitting procedure, accounting for the dust size distribution, self-consistent temperature profile and more precise radiation transfer.


\section*{Acknowledgements}
\hcadd{We thank the anonymous referee for their thoughtful and constructive comments.}
H.J. and C.W.O. would like to thank Ruobing Dong, Akimasa Kataoka, Seongjoong Kim, Sebastiaan Krijt and Chao-Chin Yang for constructive suggestions and helpful discussions during this work. The authors also appreciate feedback on an earlier version of the manuscript by Joanna Drążkowska,  Beibei Liu, Satoshi Okuzumi and Zhaohuan Zhu. This work has used \texttt{Astropy} \citep{AstropyCollaborationEtal2013}, \texttt{Matplotlib} \citep{Hunter2007}, \texttt{Numpy} \citep{HarrisEtal2020}, \texttt{Scipy} \citep{VirtanenEtal2020} software packages.

\section*{Data Availability}
The data underlying this article will be shared on reasonable requests to the corresponding author.
 



\bibliographystyle{mnras}
\bibliography{ads} 




\appendix

\section{Clump microphysical model}
\label{app:clumps}
In analogy to radiation transport, the variation of vertical mass flux $F$ with height $\mathrm{d}z$ can be simplified as two parts: the loss term $-\kappa\rho_\mathrm{c} F$ and source term $\rho_\mathrm{c}/t_\mathrm{diff}$. The first term describes the flux reduction caused by self-shielding and the second term is the new contribution due to diffusion of the local layer $\mathrm{d}z$:
\begin{equation}\label{eq:vtrans}
    \frac{\mathrm{d}\textbf{}F}{\mathrm{d}z} = -\kappa\rho_\mathrm{c} F + \frac{\rho_\mathrm{c}}{t_\mathrm{diff}}.
\end{equation}
Integrating \eq{vtrans} over the vertical dimension, we obtain
\begin{equation}\label{eq:opdiff}
    \frac{\mathrm{d} \Sigma_\mathrm{c}}{\mathrm{d} t} \equiv F = \frac{1-e^{-\tau}}{\kappa} \frac{1}{t_\mathrm{diff}} = \frac{1-e^{-\tau}}{\tau} \frac{\Sigma_\mathrm{c}}{t_\mathrm{diff}}
\end{equation}
where $\tau = \int\kappa \rho_\mathrm{c}\mathrm{d} z = \kappa \Sigma_\mathrm{c}$.
Accordingly, we define
\begin{equation}\label{eq:ftau}
  f_\mathrm{\tau} = \frac{1-e^{-\tau}}{\tau}
\end{equation}
as the diffusion suppression factor in \eq{sourcecd}. It describes how efficiently clumps can diffuse back to disc pebbles. When the "clumping depth" (optical depth for collisions) of the medium is low, replenishment of disc pebble from clumps is unopposed, this factor attains its maximum value $f_\tau\simeq1$. On the other hand, for $\tau \gg 1$, $f_\tau \ll 1$: particles in dispersing clumps do not leave the clump layer, but `bump' into other clumps.
 
In the preceding we have not yet specified the opacity $\kappa$ of the clumpy medium, which is an average over the inhomogeneous space and can therefore be lower than the geometrical opacity $\kappa_\bullet$ of the particles. For monodisperse-sized spherical particles with size $a_\bullet$, the clumping depth is defined as the collisional cross section $4 \pi a_\bullet^2$ over the single particle mass $m_\bullet$ \eq{kappadot}:
\begin{equation}
    \kappa_\bullet \equiv \frac{4\pi a_\bullet^2}{m_\bullet} = \frac{3}{\rho_\bullet a_\bullet}
\end{equation}
In terms of the Stokes number (\Eq{St}) this reads
\begin{equation}
    \kappa_\bullet = \frac{3\pi}{2}\frac{1}{\mathrm{St} \Sigma_{g}} \simeq 157\ \mathrm{cm}^2\ \mathrm{g}^{-1}\left(\frac{\mathrm{St}}{0.01}\right)^{-1}\left(\frac{\Sigma_\mathrm{g}}{3\ \mathrm{g}\ \mathrm{cm}^{-2}}\right)^{-1}.
\end{equation}
Let clumps be characterized by a radius $a_\mathrm{cl}$, volume density $\rho_\mathrm{cl}$ and let the filling factor $f_\mathrm{fill}$ be the fraction of clumps in the medium. Then
\begin{equation}
    \rho_\mathrm{cl} = \frac{1}{f_\mathrm{fill}}\rho_\mathrm{c}= \frac{Z_\mathrm{mid}}{f_\mathrm{fill}}\rho_\mathrm{g}
\end{equation}
We can calculate the clumping depth for a single clump $\tau_\mathrm{cl} = \kappa_\bullet \rho_\mathrm{cl} a_\mathrm{cl}$. With further substitution of $\kappa_\bullet$ and $\rho_\mathrm{cl}$ and assuming a vertical Gaussian distribution for gas density $\Sigma_\mathrm{g} = \sqrt{2\pi}\rho_\mathrm{g}H_\mathrm{g}$, we hence get that at a critical mid-plane dust-to-gas ratio
\begin{equation}
    Z_\mathrm{mid}^* = \frac{2}{3}\sqrt{\frac{2}{\pi}}\frac{\mathrm{St} f_\mathrm{fill} H_\mathrm{g}}{a_\mathrm{cl}}
\end{equation}
the clumps reach $\tau_\mathrm{cl} = 1$. Below this value, dust within an individual clump can easily escape and the transport of dust performs the same as single-particle disc pebbles. Thus we could keep the opacity of single dust particles as the medium opacity. On the other hand, for collisionally thick clumps, clumps act as the target particles. In that situation, denser clumps (higher $Z$) no longer increase the collision rate. We therefore write for the net opacity:
\begin{equation}\label{eq:kappa}
    \kappa(Z_\mathrm{mid}) = 
    \begin{cases}
        \frac{3}{4}\frac{1}{\rho_\bullet a_\bullet} = \kappa_\bullet & \tau_\mathrm{cl} \ll1\\
        \frac{3}{4}\frac{1}{\rho_\mathrm{cl} a_\mathrm{cl}} = \kappa_\bullet \left(\frac{Z_\mathrm{mid}^*}{Z_\mathrm{mid}}\right) & \tau_\mathrm{cl} \gg1
    \end{cases}
\end{equation}
This implies that the \emph{medium} clumping depth $\tau = \kappa\Sigma_\mathrm{c}$ will also hit a maximum:
\begin{equation}
    \tau = \min (\kappa_\bullet \Sigma_\mathrm{c}, \tau_\mathrm{max})
\end{equation}
with
\begin{equation}
    \tau_\mathrm{max}
    = \kappa_\bullet \frac{Z^\ast_\mathrm{mid}}{Z_\mathrm{mid}} \Sigma_\mathrm{c}
    = \sqrt{2\pi} \frac{ f_\mathrm{fill}H_\mathrm{c}}{a_\mathrm{cl}}
\end{equation}
where the max clumping depth $\tau$ is independent of the dense clump density. Physically, the upper limit of $\tau$ describes that only dust at the surface of clump layer ($\tau\lesssim\tau_\mathrm{max}$) can diffuse out. The value of $\tau_\mathrm{max}$ is determined by several uncertain clump properties. In general, bigger clumps that fill a smaller fraction of the space will result in a more transparent medium, and therefore result in a lower $\tau_\mathrm{max}$. Study of microphysics on the properties of clumps would be necessary to model $\tau_\mathrm{max}$, which is beyond the scope of our work. Here,  we simply parametrize this uncertainty in terms of $\tau_\mathrm{max}$, for which we take $\tau_\mathrm{max} = 15$ in our simulations. A larger $\tau_\mathrm{max}$ means that the clump sublayer will be even more clumpy for dust to escape. Particles are more trapped in the ring by collision with each other. A lower $\tau_\mathrm{max}\lesssim10$ will inversely limit the the clump layer more transparent for clump diffusing out, which will hinder the ring's survival. The conclusions of this paper are unaffected as long as $\tau_\mathrm{max}\gg1$.


\bsp	
\label{lastpage}
\end{document}